\PassOptionsToPackage{x11names}{xcolor}
\documentclass[submission, Phys]{SciPost}

\usepackage{mydefs}
\hypersetup{colorlinks=true}

\begin{document}

\fancyfoot[C]{}

\begin{center}{\Large \textbf{
Robust Independent Validation of Experiment and Theory:\\[0.2ex] \rivet version 3
}}\end{center}

\begin{center}
C.~Bierlich\textsuperscript{1,2},
A.~Buckley\textsuperscript{3},
J.~M.~Butterworth\textsuperscript{4},
C.~H.~Christensen\textsuperscript{1},
L.~Corpe\textsuperscript{4},\\
D.~Grellscheid\textsuperscript{5},
J.~F.~Grosse-Oetringhaus\textsuperscript{6},
C.~G\"utschow\textsuperscript{4},
P.~Karczmarczyk\textsuperscript{6,7},
J.~Klein\textsuperscript{6,8},
L.~L\"onnblad\textsuperscript{2},
C.~S.~Pollard\textsuperscript{9},
P.~Richardson\textsuperscript{6,10},
H.~Schulz\textsuperscript{11},
F.~Siegert\textsuperscript{12}
% ...
\end{center}

\vspace{1ex}

\begin{center}
\textbf{1} Niels Bohr Institute, Copenhagen, Denmark\\
\textbf{2} Department of Astronomy \& Theoretical Physics, Lund University, Lund, Sweden\\
\textbf{3} School of Physics \& Astronomy, University of Glasgow, Glasgow, UK\\
\textbf{4} Department of Physics \& Astronomy, University College London, London, UK\\
\textbf{5} Department of Informatics, University of Bergen, Bergen, Norway\\
\textbf{6} CERN, Meyrin, Switzerland\\
\textbf{7} Faculty of Physics, Warsaw University of Technology, Warszawa, Poland\\
\textbf{8} Istituto Nazionale di Fisica Nucleare (INFN), Torino, Italy\\
\textbf{9} DESY, Hamburg, Germany\\
\textbf{10} Institute for Particle Physics Phenomenology, Durham University, Durham, UK\\
\textbf{11} Fermilab, Batavia IL, USA\\
\textbf{12} TU Dresden, Dresden, Germany\\[1ex]
\textsmaller{* andy.buckley@cern.ch}
\end{center}


\begin{center}
\today
\end{center}



\clearpage
\fancyfoot[C]{\thepage}


\section*{Abstract}
\textbf{%
  First released in 2010, the \rivet library forms an important repository for
  analysis code, facilitating comparisons between measurements of the final
  state in particle collisions and theoretical calculations of those final
  states. We give an overview of \rivet's current design and implementation, its
  uptake for analysis preservation and physics results, and summarise recent
  developments including propagation of MC systematic-uncertainty weights,
  heavy-ion and $ep$ physics, and systems for detector emulation. In addition,
  we provide a short user guide that supplements and updates the \rivet user
  manual.%
}

\vspace{2em}
\begin{spacing}{0.92}
  \noindent\rule{\textwidth}{1pt}
  \smaller
  \tableofcontents
  \thispagestyle{fancy}
  \noindent\rule{\textwidth}{1pt}
\end{spacing}

\clearpage
\section{Overview}

Experiments at particle colliders provide many measurements of the final state
in particle collisions. These measurements range from relatively simple counts
of final state particles, to cross-sections for the production of complicated
final states multiply-differential in the kinematics of more complex objects
such as hadronic event shapes or missing energy. These measurements are
typically made in so-called ``fiducial'' regions --- that is, within a region of
phase space defined by kinematic cuts to reflect regions in which the particle
detectors have high acceptance and efficiency, thus minimising model dependence,
since large theory-based extrapolation into unobserved regions is not required.

Relatively small ``unfolding'' corrections are then often applied to account for
residual instrumental effects to within some evaluated uncertainty, meaning that
the results can be compared directly to particle-level predictions from Monte
Carlo event generators.  Unfolding is performed at the distribution rather than
event level, by constructing ``physics objects'' such as jets from physical
particles in the final state of the MC events and from there to differential
observables.  Our picture of what is a physical particle suitable for definition
of a fiducial unfolding target is nowadays usually limited to quasi-classical
colour-singlets, such as leptons direct from the hard scattering, or hadrons
(and their decay descendants) formed after the fundamental quantum dynamics have
lost coherence via non-perturbative effects.  Alternatively, a ``folding''
approach can be taken, in which the efficiency and resolution of the measuring
equipment are estimated within the measured phase space, and applied to
particle-level predictions 
to allow model-to-data comparisons.

Such measurements can contain a wealth of information about the short-distance
physics of the collision, as well as about the intervening soft processes such
as hadronisation and underlying event.  Modern theoretical calculations, within
and beyond the Standard Model, allow predictions to be made which can be
confronted with these measurements on a like-for-like basis. \rivet exists to
facilitate such comparisons, and the physics conclusions to which they lead, by
providing a set of tools to compute physical fiducial physics objects with
robust and standard definitions, and an extensive library of analysis routines
based on such definitions and immediately comparable to published data.

This document is intended to supplement and supersede the first \rivet user
manual~\cite{Buckley:2010ar}, as well as providing an overview of \rivet usage
to date and a summary of recently added features in \rivet versions up to and
including version 3.0. We first review the applications to which \rivet has been
applied, then in Section~\ref{sec:structure} review the structure to which
\rivet has evolved in its decade-long existence. In Section~\ref{sec:features}
we cover the set of major new features and functionalities since the original
paper, including the cuts system, automatic use of event weight vectors and
event groups, new mechanisms for full-accuracy run merging, tools for heavy-ion
and $ep$ physics, and tools for preservation of search analyses such as detector
emulation. We conclude in Section~\ref{sec:userguide} with a brief user guide
intended to introduce a new user to the basics of running and writing analysis
routines with \rivet.

\subsection{Applications of \rivet}

\rivet has been widely used in the development~\cite{Sjostrand:2014zea,Bellm:2015jjp,Hoeche:2012yf,Biro:2019ijx,Krauss:2018djz,Hoche:2018gti,Hoeche:2011fd},
validation~\cite{Buckley:2016bhy,CMS:2016kle,Cooper:2011gk,Heinemeyer:2013tqa,Pierog:2013ria,Ilten:2014wma,Helenius:2019gbd,Cormier:2018tog}
and tuning~\cite{Skands:2014pea,Khachatryan:2015pea,ATLAS:2012uec}
of event generators for Standard Model processes, as well as in the
study of parton density functions (PDFs)~\cite{Ball:2014uwa,Ball:2017nwa}.
Tuning generally makes use of the related Professor~\cite{Buckley:2009bj} package.

\rivet has also been used by the LHC experiments as a part of their analysis and interpretation toolkit (see, for example~\cite{Aad:2010ac,Adam:2015qaa,Aad:2015eia,Khachatryan:2014uva,Sirunyan:2018ptc}), and in studies for future experiments~\cite{Azzi:2019yne,Nejad:2016bci,Chala:2018qdf,Bothmann:2016loj}. It has been used for development
of new analysis techniques including machine learning applications, jet substructure, boosted-particle tagging and pile-up
suppression~\cite{Kasieczka:2019dbj,Monk:2018clo,Moore:2018lsr,Altheimer:2013yza}.
Extraction of SM parameters for example using TopFitter~\cite{Brown:2019pzx,Buckley:2015nca} and other phenomenological studies of the
SM~\cite{Neill:2018wtk,Reyer:2019obz,Bothmann:2018trh,Bendavid:2018nar,deFlorian:2016spz} have used \rivet, and it has also been
employed in searching for and constraining BSM physics~\cite{Papucci:2011wy,Amrith:2018yfb,Brooijmans:2018xbu,Butterworth:2019iff}, sometimes making use of the
related \contur package~\cite{Butterworth:2016sqg}.

The above list of references is incomplete, but serves to illustrate the wide applicability of, and demand for, \rivet functionality.

\section{Structure and design}\label{sec:structure}

\rivet is structured in a layered fashion, with a \cxx shared library at its
core, supplemented by \cxx ``plugin'' libraries containing collider analysis
routines, a \python programming interface built via the Cython system, and
finally a set of \python and shell scripts to provide a command-line
interface. The principle deployment targets are Unix-like systems, primarily Linux
and Mac~OS. \rivet's design is motivated by ease of use, in particular aiming to
provide a natural \& expressive analysis-writing interface with minimal
technical ``boilerplate'', as far as possible while also being computationally
efficient and supporting a wide range of use-cases.

\subsection*{Dependencies}
The core library provides machinery for structuring ``runs'' of the code,
i.e.~the feeding of simulated collider events into it for analysis, and for
output of histogram data. These input and output roles are not played entirely
by \rivet itself: it uses the \hepmc~\cite{Dobbs:2001ck,Buckley:2019xhk} and
\yoda libraries for I/O and in-memory
representation of events and histograms/analysis summary data.  \hepmc events
read into \rivet are wrapped into a more convenient |Rivet::Event| object, with
a potential event-graph tidying step before storage as the event currently being
analysed. The \yoda library was developed primarily for use with \rivet, and has
a similar layered structure with \cxx and \python interfaces and user scripts,
but is a general-purpose tool for 
statistics without particle-physics specialisations.

\subsection*{Event loop}
\begin{sidewaysfigure}
  \centering
  \includegraphics[width=0.9\textwidth]{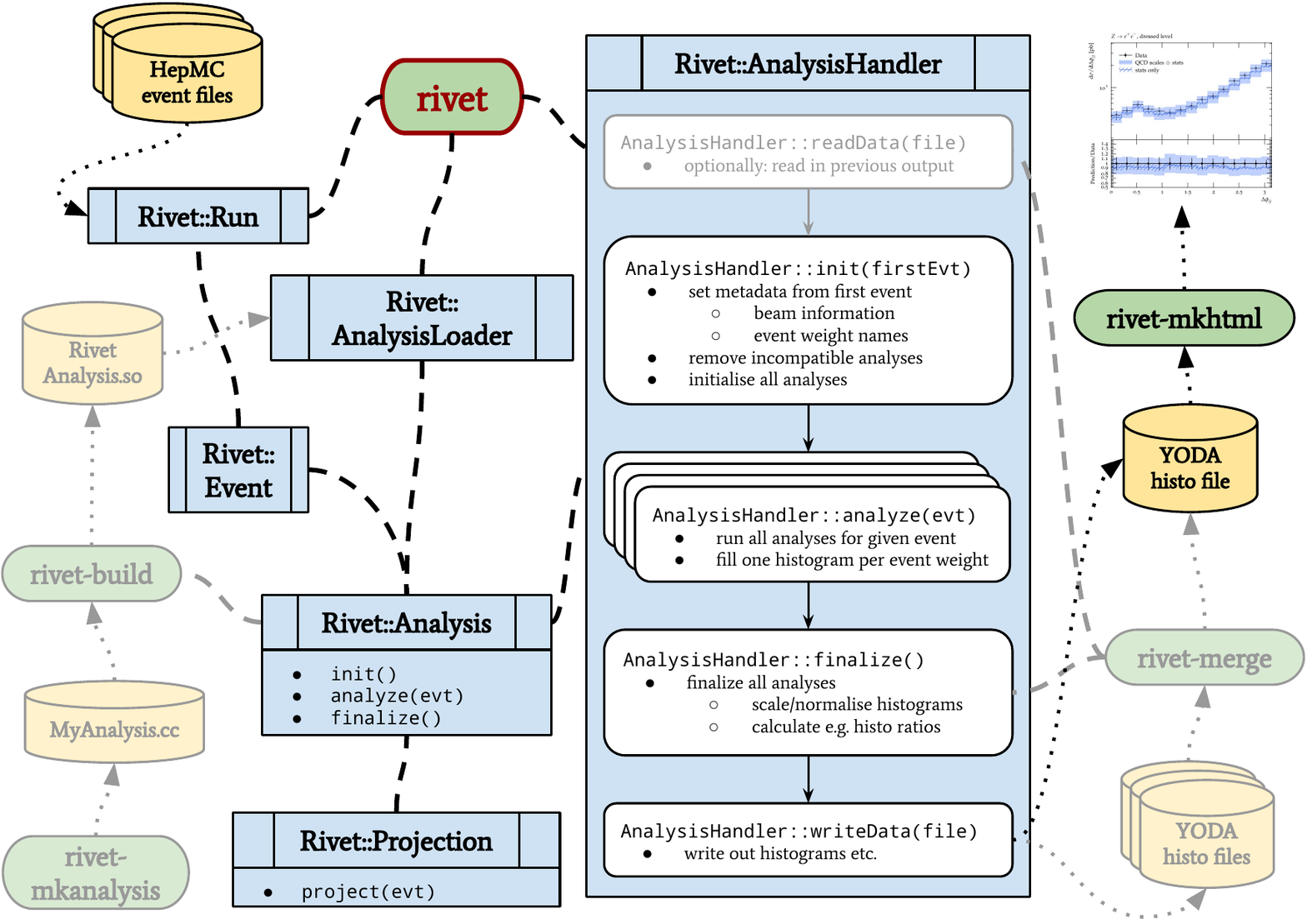}
  \caption{The \rivet system's main components and data flows. Library classes
    are shown in blue rectangles with functions in white, filesystem data in
    yellow cylinders, and user-interface scripts in green ovals. Dotted arrows
    indicate data flows, and dashed lines the main relationships between classes
    and scripts. Greyed components are optional paths for analysis preparation
    and post-processing.}
  \label{fig:rivet}
\end{sidewaysfigure}

Internally, \rivet is primarily a framework for executing analysis routines on
the incoming stream of events. The top-level structure of this framework in
terms of code objects, user-facing scripts, and data flows is illustrated in
Figure~\ref{fig:rivet}.

As with most such frameworks, this is broken into three
phases: initialisation, execution, and finalisation. \rivet analysis objects,
which inherit from the |Rivet::Analysis| base class, have three methods
(functions), which are called at each stage: |init()| during initialisation,
|analyze(const Rivet::Event&)| for each event during the execution loop, and an
optional |finalize()| called once at the end of the run.  Initialisation and
finalisation are used for the set-up and pull-down phases of the analysis, most
notably for creating histogram objects with appropriate types and binnings
during initialisation using the |Analysis::book()| methods, and scaling or
normalising them (or performing arbitrarily more complicated post-processing
manipulations and combinations) in the |finalize()| step. The execution step for
each event involves computing physical quantities from the provided event, and
using them to make control-flow decisions, and filling of histograms. However,
experience shows that many calculations are common to the majority of analyses
at a particular experiment or even whole collider: repeating such computations
for each analysis does not scale well --- as noted by the large experimental
collaborations, which centralise the processing and calculation of common
physics objects for both collider data and MC simulations. \rivet solves this
problem by a semi-opaque method named ``projections''.

\subsection*{Projections}
A projection is a stateful code object inheriting from the |Rivet::Projection|
class, which provides a |project(const Rivet::Event&)| call signature. This
operation computes a set of physical observables, e.g.~the set of all
final-state particles meeting certain kinematic or particle-ID requirements, an
event-shape (e.g.~eigenvalues of the sphericity tensor
family~\cite{Bjorken:1969wi}), or collections of collimated particle-jets. Each
projection subclass adds custom functions by which these observables may be
retrieved after computation. The strength of using projections as computational
objects rather than basic functions, is that they can be stored: this permits
caching, whereby a second (and third, and so-on) calculation of the same
quantity from the same event automatically takes the short-cut of returning the
previously calculated version.

In \rivet, this caching requires central storage of each projection, and a
mechanism for determining whether two projections are equivalent. The latter is
performed via a |SomeProj::compare(const Projection&)| method, specific to each
projection type, and the former requires that projections are registered with
the \rivet core in the initialisation phase of analysis execution. The
|compare()| method typically compares numerical configuration parameters for the
operations to be performed by the projection, e.g.~the \pt and \eta cut values,
or the clustering measure, radius, and grooming operations to be performed in
jet reconstruction. It is guaranteed that projection comparisons are only ever
performed between projection objects of the exact same type, via the \cxx
runtime type information (RTTI) system.

Registration is performed during initialisation by each analysis, using the
|Analysis::declare()| method: this compares a configured projection object to
all those currently known to the system, and either registers a \emph{clone} of
the new, distinct object in the global system with a character-string name
specific to the registering analysis, or identifies an exactly equivalent
projection and makes a new link to it via the same (name, analysis) pair. The
use of cloning during projection declaration means that memory and pointer
management problems are avoided, and the analysis-authoring user can focus on
the physics-logic of their code, rather than \cxx technicalities.

The power of projections is greatly enhanced by the ability to ``chain'' them
arbitrarily deep: any projection can itself register a set of its own
projections, like standard building blocks to construct a well-behaved
calculation. The same declaration and cloning abilities are available to
projections, with the only additional requirement on a projection-authoring user
being that they include calls to their contained projections' |compare()|
methods via the convenience |pcmp()| method.

By the end of initialisation, all projections should be declared and uniquely
registered to the \rivet core, and are known to each analysis (or chaining
projection) by a string, e.g.~|"Jets"|, or |"CentralMuons"|, or |"Thrust"|. They
may then be called by that string name using the projection/analysis'
|apply<T>()| method. This is unfortunately complicated by \cxx's unawareness of
the exact type, so the user must specify the projection type |T| that they want
the returned object to be. The |auto| keyword in \cxx{11} makes this operation
slightly simpler, and overall the configuration and calling of projections is a
simple operation that automatically provides result-caching for common
quantities, as well as encapsulating a lot of standard detail and workarounds
for misconceptions and problems in computation of physical quantities from
\hepmc/\rivet events.

\subsection*{Event analysis}
During the execution phase, both analyses and projections can also use more
direct physics calculation tools. The most important of these are the
|Particle|, |Jet| and |FourMomentum| classes, and an array of functions for
computing kinematics or particle ID properties, e.g.~a set of |deltaR()|,
|mT()|, |isCharged()|, |hasCharm| etc.\ functions which are also largely
replicated on the class interfaces of |Particle|, |Jet| and |FourMomentum|
themselves. From \rivet~2.6 onwards, the |Particle| class acquired the ability
to recursively contain more |Particle|s, useful for definition of reconstructed
``pseudoparticle'' objects like $Z$ bosons, reconstructed top-quarks, and
photon-dressed charged leptons. This development also brings particles and jets
conceptually closer together and indeed many useful functions (e.g.~ancestor and
descendent checks) are provided for general application to their common base
class, |ParticleBase|. The two concepts are not yet fully unified, however, with
|Particle| but not |Jet| providing particle-ID information, and |Jet| but not
|Particle| supporting truth-tagging using ghost-associated $c$ and $b$ hadrons,
and $\tau$ leptons.

Particles and jets are returned by their respective ``finder'' projections as
lists, known as |Particles| and |Jets| containers, which are often sorted in
decreasing \pt. Picking the leading objects from such vectors is hence easy
using standard vector indexing, but sometimes more complex operations are
needed, such as filtering out objects that do or don't meet certain criteria, or
summing the \pt{}s of all the objects in a list to compute an \HT or \meff
measure. \rivet makes use of modern ``functional'' programming capabilities in
\cxx for this, in the form of |select()| and |discard()| functions, which take a
list and a ``functor'' (function object) as arguments: the function then returns
a new list containing the physics objects from the first list which caused the
functor to return true or false, depending on whether the logic is to have
selected or discarded those which returned true. This system is immensely
flexible and far more convenient than standard-library tools, and will be
described later in more detail, with examples. Particle and jet lists may also
be concatenated using the |+| operator, unlike STL vectors.

\subsection*{Analysis-routine compilation \& loading}
Analyses are compiled separately from the core \rivet library, as ``plugin''
shared libraries which are loaded explicitly by searching the internal analysis
path (set either programmatically or via the |RIVET_ANALYSIS_PATH| environment
variable) for libraries matching the pattern |Rivet*.so|. A |rivet-build| script
is supplied as a plugin-building frontend for the \cxx compiler, both for
simplicity and to ensure exact compatibility of compiler options with the core
library. This script can build multiple analyses into a single plugin, and as
well as for user convenience is used to build the more than 700 included
analysis codes into plugins grouped by experiment or collider. 


\section{New features}\label{sec:features}

\subsection{Combineable kinematic cuts and filtering functors}\label{sec:cuts}

Clarity and expressiveness of analysis logic are key to the \rivet design
philosophy, and continuous development of the \rivet API has refined how this
goal has been technically achieved. Ideal demonstrations of this have been the
addition of major code systems for more configurable expression of kinematic
cuts and other physics-object filtering, as well as cosmetic reduction of
clutter such as
\begin{itemize}
\item replacement of the rather cumbersome |addProjection()| and
  |applyProjection()| functions with neater |declare()| and |apply()| names;
\item direct provision of momentum-like properties on |Particle| and |Jet| types
  without needing to retrieve the contained |FourMomentum|;
\item automatic implicit casting of |Jet| and |Particle| to |FourMomentum| to
  functions expecting the latter as an argument, and similar implicit casting between \fastjet and \rivet jet types; and
\item provision of |abseta()|, |absrap()| and |abspid()| methods on physics
  objects to reduce the parenthetic noise introduced by user calls to |abs()| or |fabs()|.
\end{itemize}
Here we summarise first the |Cuts| system, and then its extension to filtering
functors.

In the original API, many functions used
lists of several, perhaps optional, floating-point valued arguments,
e.g.~|FinalState(double etamin, double etamax, double ptmin)| or
|fs.particles(double ptmin)|. Not only were these inflexible, not allowing
``inline'' cuts other than the hard-coded ones, and sometimes annoyingly verbose
--- as with nearly always symmetric rapidity cuts, or often having to supply
|DBL_MAX| values to apply a \pt cut only --- but at the compiled code level they
were ambiguous. It was easy to forget if the $\eta$ cuts or the \pt cut came
first, and accidentally require something like $\eta > 10$, with clear
consequences: the compiler has no way of telling one |double| from another and
warning the analysis author of their error.

To address this without adding surprising behaviours, myriad specially
named functions, or other undue complexity to analysis code, we developed a
|Cut| object based on \cxx{11} smart pointers, with function overloads allowing
|Cut|s to be combined using any normal boolean operators, e.g.~|operator(const Cut&, const Cut&)| %
$\to$ |Cut|. Several specialisations of |Cut| are provided in
the \kbd{Rivet/Tools/Cuts.hh} header, providing |enum|s which map to cuts on \pt,
rapidity, pseudorapidity, energy, \et, charge, etc. These can be applied on
|FourMomentum|, |Particle| and |Jet| objects, and for |Particle| a |Cuts::pid|
|enum| is available in addition, supporting equality as well as inequality
comparisons and for use with particle-ID |enum|s like |PID::ELECTRON|. This PID
cut, as well as the rapidities and charges, is also available in ``abs'' form
i.e.~|Cuts::abspid|, |Cuts::abseta|, etc., to allow clearer and more compact
analysis cut expressions. The use of explicit cut expressions like
|fj.jets(Cuts::pT > 20*GeV && Cuts::absrap < 2.5)| has made \rivet analysis code
both easier to write and to read.

However, these cuts are still static, predefined entities: useful for the
majority of object selection requirements, but not all. In particular, the cuts
are defined for one object at a time, while an important class of analysis cuts
select or reject objects depending on their relations to other objects in the
event, e.g.~lepton--jet isolation or overlap-removal. The logic of such
isolations is not particularly complex, but typically involves several nested
|for|-loops, which require reverse-engineering by a reader wishing to understand
the analysis logic. In addition, the \cxx Standard Template Library interface
for deleting objects from containers --- the so-called ``erase--remove idiom''
--- is verbose and non-obvious, distracting from the physics goal of object
filtering. For this reason, the |(i)filter| functions already described were
added, as well as event higher-level functions like the |(i)select/discardIfAny|
set, which accept \emph{two} containers and a comparison functor, for
e.g.~implicitly looping over all leptons and in-place discarding any jet that
overlaps with any of them.

These have revolutionised the writing of isolation-type code, reducing
complex multi-loop code to one-line calls to a filtering function, leveraging
the substantial library of function objects. Use of the \cxx{11} |std::function|
interface means that these functors can be normal functions, e.g.~the
|isCharged(const Particle&)| utility function, but powerfully they may also be
stateful objects to implement configurable cuts such as |pTGtr(const FourMomentum&, double)|, %
or |hasBTag(const Cut&)|. This latter class allows the
function to be defined with respect to another object, selected ``live'' in the
body of a loop, and even \emph{defined} inline, using the \cxx{11} anonymous
``lambda-function'' system.

Similar uses of the functional coding paradigm in \cxx{11} have led to other
classes of functor for sorting (which may be passed to functions like
|ParticleFinder::particles()| or |JetFinder::jets()|), and for computation of
variables (for use with functions such as |sum(const Particles&)|). Again, these
can accept functions or function objects, including \cxx inline lambda
functions.  With care, such interface evolutions have provided a step-change in
the power and expressiveness (and compactness) of \rivet analysis logic, not
just without sacrificing readability and self-documentation, but improving it.

\subsection{Systematic event weights}
\label{sec:weights}

One of the main structural changes in \rivet~3 is the handling of event
weights. In the quest of understanding better the uncertainties in the
event generator modelling, it is important to vary scales, parameters,
parton density functions, etc. Previously, this was handled by
generating many event samples corresponding to different
settings. Today most event generators instead accomplish the same
thing by assigning a set of different weights for each generated
event corresponding e.g.~to different scale choices.

Another aspect of weights is that in some next-to-leading order (NLO)
QCD generators there is a system of combining events into event
groups. In e.g.\ a dipole subtraction scheme \cite{Catani:1996jh},
real emission events would be accompanied by a set of counter events
with Born level kinematics corresponding to the possible dipole
mappings of the real emission phase space. The real emission event
would be weighted according to the tree-level cross section, while the
counter events would be negatively weighted according to the
corresponding Born cross section times the releavnt dipole splitting.
This means that great care must be taken when analysing these events
and filling a histogram with the weights. For well behaved (soft and
collinearly safe) observables the real and counter events will
normally end up in the same bin. But due to the different underlying
kinematics it may happen that they don't. Also, to obtain the correct
statistical uncertainty, a histogram bin cannot be filled once with
each (sub-)event. Instead, the bin should be filled once for each such
\textit{event group}, with a weight given by the sum of the weights of
the real and counter-events.

Finally, matching the ever increasing luminosity of today's colliders
often requires the generation of extremely large event samples. To make
this efficient, the generation is often divided in to many parallel
runs that need to be combined afterwards, treating the weights in a
statistically correct way. To make sure that the weights are treated correctly in all
such cases, the weight handling is no longer directly exposed to the
implementer of an analysis, but is handled behind the scenes, as described
in the following.

\subsubsection{Handling of multiple event weights}
\label{sec:handl-mult-weights}

To avoid having to run each analysis many times for the same event,
i.e.~once for each supplied event weight, the weight handling is no
longer directly exposed to the |Analysis| class. This means  that the
histograms and other analysis objects to be filled in an analysis are
encapsulated in a wrapper class actually containing several copies of
the same histogram, one for each event weight. For the analysis
implementer things look more or less as in \rivet~2. The histograms
are still handled with pointers, |Histo1DPtr|, but while before these were
standard shared pointers, they are now much more sophisticated.

The main visible changes are in the booking and the filling of
histograms. While before a histogram would be booked as %
|hist = bookHisto1D(...)|, the syntax has changed to %
|book(hist, ...)| (and similarly for the other analysis object
types). In addition, rather than always having to explicitly fill a histogram
with a weight in \rivet~2 as \\
|hist->fill(x,weight)|, the new scheme will handle the weights
separately changing this syntax to\footnote{A second argument can still be supplied here, but this is then interpreted as a number that should be multiplied by the event weight} |hist->fill(x)|.

What happens behind the scenes is that |book(...)| will actually
create several \yoda histograms, one for each weight. This means that it
is not possible at this point in the |Analysis::init()| function to
actually access a particular instance of the histogram through the
|Histo1DPtr| pointer. In the same way in the |Analysis::analyze(...)|
function it is only possible to fill the histograms using the
|fill(...)| function, while any other attempt to manipulate a
particular histogram will fail. Calling the |fill(...)| function will
actually not directly fill each histogram with the corresponding event
weight, rather the fills will be recorded in a list, and the actual
histogram filling will only take place after all analysis of an event
(group) has been completed. The reason for this will become clear in
Section~\ref{sec:handling-nlo-events} below.

At the end of a run, |Analysis::finalize()| is called for all
analyses, once for each event weight. The syntax in this function is
completely unchanged from \rivet~2, and in each call the |Histo1DPtr|
will work as a normal pointer to a |YODA::Hist1D| object, that can be
manipulated in the normal way.

It is worth noting that the implementer typically does not need to worry
about the event weight when writing code.
This of course assumes that the user is not expected
to fill a histogram with a combination of different event weights.
Such an event-weight manipulation is better handled within the actual
generators, and hence the generators are expected to produce
self-contained event weights ready for histogram filling, with the
exception of counter-events discussed in the next section.

The weights are taken from the input \hepmc file, where they must be
given a name. There is so far no standard format for the weight names,
and basically any character string can be handled by \rivet. It should
be noted, however, that names of analysis objects in the output \yoda
file will have the corresponding weight names appended, enclosed in
square brackets, and in general it is not advisable to use
special characters in the weight names. In addition \rivet will treat
one of the weight supplied in the \hepmc as \textit{nominal}, and the
corresponding analysis objects will be stored in the output \yoda file
without the weight name appended. Also for the nominal weight, there
is no fixed convention for the name, and \rivet will assume that a
weight named |""| (empty string), |"0"|, |"Default"| or |"Weight"| is
the nominal one. If there is no weight with such a name, the first
weight found will be treated as nominal.

Handlers are provided in the \python interface to extract or strip
the weight names from histogram paths.

\subsubsection{Handling of NLO events and counter-events}
\label{sec:handling-nlo-events}

When handling an event group of NLO-generated real emission events
with corresponding counter-events, the events are fully correlated and
it is important that each event group is treated as one, so that each
histogram fill is given by the sum of the event weights in the group. In
addition these fills should have the correct error propagation encoded
in the sum of weights (\sow) and the sum of the squared weights (\sosw).

The idea is that in a histogram of a soft- and collinear-safe
observable, the real emission event and the corresponding counter-events
will always end up in the same bin in the limit when the real
emission is soft and/or collinear. In this limit, the weight for the
real event approaches positive infinity and the weight of one or more
of the counter-events approaches negative infinity. However, there is
always a possibility that a fill of the real event ends up very close
to a bin edge while a counter-event ends up on the other side of the
edge, ruining the NLO cancellation. In \rivet the trick to solve this
problem is to not just fill at the given value of the observable, but
to spread it out in window, possibly filling also adjacent bins.

The full procedure to handle this with correct error propagation is
fairly complicated, and is completely hidden from the user and
analysis implementer. For reference the full description if the
procedure is given in Appendix~\ref{sec:handl-groups-compl}.

\subsubsection{Event-weight analysis}
\label{sec:weight-analysis}

In recent years it has become increasingly important to study the
event-weight distribution directly. Producing very large event samples
can quickly become an expensive endeavour when the overall sample
size is determined by the delicate trade-off between the desired statistical
precision and how much of the CPU budget is being spent simulating events
with negative event weights, which would ultimately reduce the
statistical power of the sample.
For unweighted events, both the spread of the event weights
as well as the fraction of negative weights in the
sample need to be understood in order to be able to project the CPU cost
of the sample correctly. Although \rivet~3 has put a lot
of effort into hiding the handling of the event weights from the user,
it is still possible to retrieve them and treat them as an observable
when filling a histogram.
The corresponding syntax is illustrated in the routine |MC_XS|.
Note that this will not work if the sample makes use of counter-events.

\subsubsection{Re-entrant finalize and run merging}
\label{sec:reentrant-finalize}

It is quite common to divide up \rivet analyses of very large event
samples in smaller runs, and merge the produced \yoda
files. Previously this has been done with the |yodamerge| \python
script distributed with \yoda. However, this script was completely
ignorant of the way the different analysis objects were produced.
In order to better handle this \rivet has since version~2.7 introduced
the \python script |rivet-merge| which does the same thing but using the
complete knowledge of how the analysis objects were filled in
|Analysis::analyze(...)| and manipulated in |Analysis::finalize()|.

The way this is implemented means that all analysis objects come in two
different instances. One is called \textit{raw} and is used
only for filling in the |Analysis::analyze(...)| functions. Before a
call to |Analysis::finalize()| the raw analysis objects are copied to
the \textit{final} instances which are then used for manipulating the
objects into their final form to be plotted. In this way the
|Analysis::finalize()| function can be run several times and is then
called \textit{re-entrant}. The user will notice that the output \yoda
file contains both sets of analysis objects, one with the standard
naming, and one with the same name prefixed by |/RAW/|.
\footnote{Handlers are provided in the \python interface to identify these histograms
when reading \yoda files for post-processing.}

In this way the |rivet-merge| script can read in \yoda files, create
and run the |Analysis::init()| for the corresponding analyses, merge
all raw analysis objects together and run |Analysis::finalize()|. When
doing this it is important to note that there are two different kinds
of merging possible. In one situation we have run \rivet on several
completely equivalent event samples, and in the other the runs have
been on different kinds of event samples which should be combined in
another way. An example of the latter is when the same analysis is run
on event samples with different $\sqrt{s}$, and in the end
will produce e.g.~ratios of distributions between different energies.

To understand the differences between these two merging schemes we assume we
have a number of runs, each with the cross-section $\sigma_i$ as reported by the
generator, the total sum of weights $S_{w,i}$, and a number of ``raw'' histogram
bins each with a sum of weight $S_{w,i}^{[b]}$ and the sum of squared weights
$S_{w^2,i}^{[b]}$. When finalized the histogram bin will typically have a
cross-section $\sigma_i^{[b]}=\sigma_iS_{w,i}^{[b]}/S_{w,i}$ with an estimated
error of $\delta\sigma_i^{[b]}=\sigma_i\sqrt{S_{w^2,i}^{[b]}}/S_{w,i}$. But we
can also imagine other situations where the final plot point is a fraction of
the total $r_i^{[b]}=\sigma_i^{[b]}/\sigma_i$, or if we want to be a bit more
general any ratio of cross-sections
$r_i^{[b/a]}=\sigma_i^{[b]}/\sigma_i^{[a]}$. We note that for uniformly weighted
events an individual generator run corresponds to an integrated luminosity
${\cal L}_i = S_{w,i}/\sigma_i$.

Now if the runs to be merged have exactly the same process, the
weights in the combined histogram will simply be the added weights,
\begin{equation}
  S_w=\sum_i S_{w,i},\quad S_w^{[b]}=\sum_i S_{w,i}^{[b]}, \quad\mathrm{and}
  \quad S_{w^2}^{[b]}=\sum_i S_{w^2,i}^{[b]}
\end{equation}
and the cross-section for the combined files will be a weighted average
\begin{eqnarray*}
  \sigma &=& \frac{1}{S_w} \sum_iS_{w,i}\sigma_i \, .
\end{eqnarray*}

Alternatively, to be on the safe side, if the files have identical
processes, but different weight variations, we might want to use the
effective number of entries, $\mathcal{N}=S^2_w/S_{w^2}$, as weights,
\begin{equation}
  \sigma = \frac{1}{\mathcal{N}} \sum_i\mathcal{N}_i\sigma_i\quad\mathrm{and}
  \quad\delta\sigma^2=\frac{1}{\mathcal{N}^2}\sum_i\mathcal{N}_i^2\delta\sigma_i^2 \, .
\end{equation}
For each bin we will e.g.~have the plot value
$\sigma^{[b]}=\sigma S_{w}^{[b]}/S_{w}$ with an estimated error of
$\delta\sigma^{[b]}=\sigma\sqrt{S_{w^2}^{[b]}}/S_{w}$ as for the
individual histograms.

Turning now to the case of adding histograms with different processes,
the case where the histograms are already normalised to cross-section
is the easiest, since we can then simply add
\begin{equation}
  \sigma=\sum_i\sigma_i,\quad\sigma^{[b]}=\sum_i\sigma_i^{[b]},\quad\mathrm{and}
  \quad\delta\sigma^{[b]}=\sqrt{\sum_i\left(\delta\sigma_i^{[b]}\right)^2} \, .
\end{equation}

For adding the raw histograms we need to expand out the cross-sections
in terms of weights,
\begin{eqnarray*}
  \sigma\frac{S_w^{[b]}}{S_w}=\sum_i\sigma_i\frac{S_{w,i}^{[b]}}{S_{w,i}},\quad
  \mathrm{and}\quad\delta\sigma^2\frac{S_{w^2}^{[b]}}{S^2_w}=\sum_i\frac{\delta\sigma_i^2S_{w^2,i}^{[b]}}{S_{w,i}^2} \, .
\end{eqnarray*}
In other words, the ratio of the weights to the total is a
cross-section-weighted average, and we can write
\begin{equation}
  S_w^{[b]}=\frac{S_w}{\sigma}\sum_i\sigma_i\frac{S_{w,i}^{[b]}}{S_{w,i}}
  \quad\mathrm{and}\quad
  S_{w^2}^{[b]}=\frac{S_w^2}{\sigma^2}\sum_i\frac{\sigma_i^2S_{w^2,i}^{[b]}}{S_{w,i}^2} \, .
\end{equation}

However, the $S_w$ is arbitrary (two equations and three unknowns
above), and this is related to the fact that the combined histograms
no longer necessarily corresponds to a particular integrated
luminosity. This, in turn, means that it is not possible to first
combine histograms of different processes and then combine these with
others of identical combinations.

If the different runs do correspond to the same integrated luminosity,
of course the combined run should correspond to the same. One
reasonable way of obtaining this could be to let the integrated
luminosity for the merged sample be the cross-section-weighted average
of the individual samples,
\begin{equation}
  \frac{S_w}{\sigma}={\cal L}=
  \frac{1}{\sigma}\sum_i\sigma_i{\cal L}_i=\frac{\sum_i S_{w,i}}{\sigma} \, .
\end{equation}

In conclusion, the way |rivet-merge| combines raw histograms for
different runs is for different processes (the default)
\begin{equation}
  S_w^{[b]}=\frac{S_w}{\sigma}\sum_i\sigma_i\frac{S_{w,i}^{[b]}}{S_{w,i}}
  \quad\mathrm{and}\quad
  S_{w^2}^{[b]}=\frac{S_w^2}{\sigma^2}\sum_i\frac{\sigma_i^2S_{w^2,i}^{[b]}}{S_{w,i}^2},
\end{equation}
while for combining histograms with identical processes (using the
command argument |-e| or |--equiv|)
\begin{equation}
  S_w^{[b]}=\sum_i S_{w,i}^{[b]}\quad\mathrm{and}\quad
  S_{w^2}^{[b]}=\sum_i S_{w^2,i}^{[b]} \, .
\end{equation}
Similarly the resulting \yoda file will have
\begin{equation}
  S_w=\sum_i S_{w,i}\quad\mathrm{and}\quad S_w^2=\sum_i S^2_{w,i}
\end{equation}
in all cases, while for identical processes
\begin{equation}
\sigma=\frac{1}{\mathcal{N}}\sum_i\mathcal{N}_i\sigma_i\quad\mathrm{and}\quad
\delta\sigma^2=\frac{1}{\mathcal{N}^2}\sum_i\mathcal{N}_i^2\delta\sigma_i^2 ,
\end{equation}
and for different processes
\begin{equation}
  \sigma=\sum_i\sigma_i\quad\mathrm{and}\quad
  \delta\sigma^2=\sum_i\delta\sigma_i^2 \, .
\end{equation}

It is important to note that not all analyses in \rivet necessarily have
re-entrant |Analysis::finalize| methods. Work is in progress to convert them
all, but it is not yet finished. The ones that have been done are given the
status |REENTRANT| in the |.info| file. 
The requirements that need to be met for this status are as follows:
\begin{itemize}\itemsep 0mm
\item All information that is needed to |finalize| an analysis
  \emph{must} be encoded in properly booked analysis objects.
\item \emph{All} analysis objects that can be used in an analysis must be
  properly booked in the |Analysis::init()|, also those that are not
  needed in the current run (e.g.~if the analysis can be run at
  different $\sqrt{s}$).
\item Non-fillable analysis objects (such as |Scatter2D|) cannot be
  merged automatically. These are normally only constructed in
  |Analysis::finalize()| as e.g\ the result of dividing two histograms, and
  it is recommended to book these analysis objects in the
  |Analysis::finalize()| method rather than in |Analysis::init()|.
\end{itemize}
An example analysis with the |REENTRANT| status is |ALICE_2012_I930312|.

\subsection{Heavy ion physics}\label{sec:hion}
%

The \rivet~framework as such, has no preference for any type of collision system,
as long as simulated collider events by a Monte Carlo event generator can be
represented in the \hepmc~format. The possibility to add experimental analysis plugins
based on data from heavy ion experiments is as such not new, and was for instance used
in the implementation of the analysis |LHCF_2016_I1385877|~\cite{Adriani:2015iwv}.

The possibility of having a heavy ion beam is, however, not sufficient to implement the
bulk of existing experimental analyses, as the employed techniques differ from standard
techniques in proton-proton and electron-positron collisions. The threshold for implementing
even simple analyses has thus previously been too high for any real progress to be made into
this area. From \rivet~2.7.1 onward, new |Projection|s and other tools to facilitate the implementation of
heavy ion analyses have been added~\cite{rivethi}, and carried through to version~3.
New features include:
\begin{itemize}
	\item A designated centrality framework, outlined in Section~\ref{sec:centrality}.
	\item A framework for calculating flow observables, based on the Generic Framework \cite{Bilandzic:2010jr,Bilandzic:2013kga}, outlined in Section~\ref{sec:generic-framework}.
	\item Designated |PrimaryParticle| projections for implementing experimental definitions of primary and secondary particles.
	\item Re-entrant finalization (see Section~\ref{sec:reentrant-finalize} for a separate introduction) to allow for
	heavy ion to $pp$ ratio figures, such as nuclear modification factors $R_{AA}$, but also useful for statistically
	correct merging in general.
	\item Pre-loading of calibration data and analysis options (see Section~\ref{sec:analysis-options} for a separate introduction) to allow for centrality selection,
	but also useful in several other cases.
	\item An |EventMixingFinalState| projection to allow Monte Carlo generated events to be mixed, to allow access to particles from distinct events in order to correct correlation functions for the effects from limited acceptance and single particle distributions.
\end{itemize}
The technical use of the centrality framework and the framework for flow observables, is outlined in Sections~\ref{sec:centrality} and~\ref{sec:generic-framework} respectively.
For a more complete overview of all new methods,
as well as a physics introduction, the reader is referred to Ref.~\cite{rivethi}.
The tools introduced for heavy-ion physics, are not limited in use to analyses of heavy-ion beams.
Already a number of analyses of $pp$ collisions implementing such techniques are made available. A full list of all
currently available analyses either implementing heavy ion functionality, or containing heavy~ion beams, is given
in Table~\ref{table:analyses}.

\begin{table}
  \begin{adjustbox}{width=\textwidth,center}
  \begin{tabular}{llllr}
	\toprule
	Analysis name & System & Validated & Heavy ion features & Reference \\
	\midrule
	\verb:ALICE_2010_I880049:  & PbPb & Yes & centrality, primary particles, & \cite{Aamodt:2010cz} \\
	\verb:ALICE_2012_I930312:  & PbPb & Yes & centrality, heavy ion container, re-entrant finalize & \cite{Aamodt:2011vg} \\
	\verb:ALICE_2012_I1127497: & PbPb & Yes & centrality, heavy ion container, re-entrant finalize & \cite{Abelev:2012hxa} \\
	\verb:ALICE_2012_I1126966: & PbPb & No & centrality, primary particles   & \cite{Abelev:2012wca} \\
	\verb:ALICE_2013_I1225979: & PbPb & No & centrality, primary particles   & \cite{Abbas:2013bpa} \\
	\verb:ALICE_2014_I1243865: & PbPb & No & centrality, primary particles   & \cite{ABELEV:2013zaa} \\
	\verb:ALICE_2014_I1244523: & $p$Pb  & No & centrality, primary particles & \cite{Abelev:2013haa} \\
	\verb:ALICE_2016_I1394676: & PbPb & No & centrality, primary particles   & \cite{Adam:2015kda} \\
	\verb:ALICE_2016_I1419244: & PbPb & No & centrality, generic framework   & \cite{Adam:2016izf} \\
	\verb:ALICE_2016_I1507090: & PbPb & No & centrality, primary particles   & \cite{Adam:2016ddh} \\
	\verb:ALICE_2016_I1507157: & $pp$   & No & event mixing                  & \cite{Adam:2016iwf} \\
	\verb:ATLAS_2015_I1360290: & PbPb & No & centrality                      & \cite{Aad:2015wga} \\
	\verb:ATLAS_2015_I1386475: & $p$Pb  & No & centrality                    & \cite{Aad:2015zza} \\
	\verb:ATLAS_PBPB_CENTRALITY: & PbPb & No & centrality                    & \cite{Aad:2015wga} \\
	\verb:ATLAS_pPb_Calib:     & $p$Pb  & No & centrality                    & \cite{Aad:2015zza} \\
	\verb:BRAHMS_2004_I647076: & AuAu & No  & centrality, primary particles  & \cite{Bearden:2004yx} \\
	\verb:CMS_2017_I1471287:   & $pp$   & No & generic framework             & \cite{Khachatryan:2016txc} \\
	\verb:LHCF_2016_I138587:   & $p$Pb  & No & ---                           & \cite{Adriani:2015iwv} \\
	\verb:STAR_2016_I1414638:  & AuAu & No & centrality                      & \cite{Adamczyk:2016exq} \\
	\bottomrule
  \end{tabular}
  \end{adjustbox}
  \caption{\label{table:analyses} All \rivet analyses implementing one or more heavy ion features, or heavy ion beams. An up-to-date list can be found at \texttt{https://rivet.hepforge.org}.}
\end{table}

\subsubsection{Centrality estimation}
\label{sec:centrality}
The size and transverse shape of the interaction region is of particular interest in the analyses of
colliding nuclei, but cannot be measured directly in experiments. Experiments instead classify
collisions according to a single event observable $N$, defining the centrality in percentiles
of the distribution $\mathrm{d}\sigma_{\mathrm{inel}}/\mathrm{d}N$, such that the centrality
of a collision is:
\begin{equation}
	\label{eq:centrality}
	c = \frac{1}{\sigma_{\mathrm{inel}}} \int_N^\infty \frac{\mathrm{d}\sigma_{\mathrm{inel}}}{\mathrm{d}N'} \mathrm{d}N'.
\end{equation}
The single event observable $N$ can then be defined in one of three ways:
\begin{itemize}
	\item As the measured distribution by the experiment, translating the percentile cuts directly to
		cuts in a measured observable. This is of course the definition most in line with the \rivet
		philosophy, but is not always feasible.
	\item In a similar way as the experiment, but using the Monte Carlo generator to generate
		$\mathrm{d}\sigma_{\mathrm{inel}}/\mathrm{d}N$, defining the percentile cuts.
	\item Using a model's impact parameter ($b$) in place of $N$, thus comparing a theoretical centrality
		to the measured one.
\end{itemize}

In experiments, $N$ is often chosen to be an observable proportional to particle production in the forward region.
Since the ability of a given Monte Carlo generator to reproduce this specific observable should not be a limiting
factor, the two latter options have been added. In such cases, the distribution $\mathrm{d}\sigma_{\mathrm{inel}}/\mathrm{d}N$
must be known before the execution loop is initiated, i.e.~when the method |analyze(const Rivet::Event&)| is
called for the first time. To that end, a calibration run using a special calibration analysis must be performed.

The calibration analysis is a simple analysis with the sole purpose of filling histograms containing the
distributions $1/\sigma_\mathrm{inel} \, \mathrm{d}\sigma_{\mathrm{inel}}/\mathrm{d}N$ and
$1/\sigma_{\mathrm{inel}} \, \mathrm{d}\sigma_{\mathrm{inel}}/\mathrm{d}b$. The output from running this analysis
is read in using the |--preload| option. This option reads the filled histogram objects into \rivet, and makes them accessible
for the duration of the (second) run. A |CentralityProjection| can then be booked by calling
|declareCentrality(const SingleValueProjection &proj, string calAnaName, string calHistName, const string projName)|.
Here |proj| is a projection returning the current value of $N$, |calAnaName| is the name of the calibration analysis,
|calHistName| is the name of the calibration histogram and |projName| is the given name of the centrality projection.
In the execution loop, the projection can be applied to the current event, and the method |cent()| will
return $c$ for the current event.

The user can select between the above mentioned centrality definitions at runtime, using an analysis option
(see Section~\ref{sec:analysis-options}). The analysis option |cent=REF| (default) selects the measured distribution,
|cent=GEN| selects the generated version of the measured distribution, |cent=IMP| the impact parameter distribution and
finally |cent=USR| allows the user to use a hard coded centrality value from the \hepmc input file (from \hepmc~3.0).

\subsubsection{Flow measurements}
\label{sec:generic-framework}
A large subset of analyses of high energy heavy ion collisions, are concerned with studies of the azimuthal anisotropy
of particle production. This is quantified in flow coefficients $v_n$'s, defined as the Fourier expansion of the particle yield
with respect to the event reaction plane $\Psi_n$:
\begin{equation}
	\label{eq:flowdefinition}
	E\frac{\mathrm{d}^3N}{\mathrm{d}^3p} = \frac{1}{2\pi}\frac{\mathrm{d}^2N}{p_\perp \mathrm{d}p_\perp \mathrm{d}y}
	\left(1 + 2 \sum_{n=1}^\infty v_n \cos([n(\phi - \Psi_n)])\right).
\end{equation}
Here $E$, $p_\perp$, $\phi$ and $y$ denote the particle energy, transverse momentum, azimuthal angle and rapidity, respectively. Since the reaction plane
is not accessible experimentally, flow coefficients are often estimated from two- or multi-particle correlations. In \rivet we have implemented the Generic Framework
formalism~\cite{Bilandzic:2010jr,Bilandzic:2013kga}, plus a number of convenient shorthands. The framework allows for quick evaluation of multi-particle correlations in terms of $Q$-vectors:
\begin{equation}
	Q_n = \sum_{k=1}^M \exp(in\phi_k),
\end{equation}
for an event of $M$ particles. Since a $Q$-vector requires just a single loop over data, as opposed to $m$ loops for an $m$-particle correlation,
the Generic Framework reduces the computational complexity of multi-particle correlation analyses from $\mathcal{O}(M^m)$ to at most $\mathcal{O}(M\log(M))$.
For a more thorough introduction to the Generic Framework, the reader is referred to the dedicated paper on heavy ion functionalities in \rivet~\cite{rivethi},
as well as Refs.~\cite{Bilandzic:2010jr,Bilandzic:2013kga}. The following text will be mainly concerned with the technical usage in analysis code.
In general, the Generic Framework expresses flow coefficients of $n$'th order in terms of $m$-particle cumulants of $n$'th order, denoted $c_n\{m\}$.
Cumulants are again expressed as correlators of even order $\langle \langle m \rangle \rangle_{n_1, n_2,...,-n_{m/2},...,-n_m}$, which can finally be
expressed algorithmically in terms of $Q$-vectors.

In order to access the Generic Framework functionality in \rivet for calculation of cumulants, the analysis must inherit from the |CumulantAnalysis|
class, which itself inherits from the |Analysis| base class. This allows for $m,n$-correlators to be booked with a call to the templated method
|template<unsigned int N, unsigned int M> bookECorrelator(const string name, const Scatter2DPtr hIn)|. Here the template arguments corresponds to
$m$ and $n$, |name| is the given name of the correlator and |hIn| should contain the binning of the correlator (usually imported from the analysis |.yoda| file).
Also available, is a |Correlators| projection, which is declared using the constructor |Correlators(const ParticleFinder& fsp, int nMaxIn, int pMaxIn)| (also available in a $p_\perp$
binned version). Here |fsp| is an already declared |ParticleFinder| derived projection from which particles should be drawn, |nMaxIn| is the maximal sum
of harmonics to be generated (e.g.~4 for $c_2\{2\}$) and |pMaxIn| the maximal number of particles to be correlated. If all desired correlators for a given
analysis are already booked, the two maximal values can be extracted automatically from booked correlators by calling |getMaxValues()|, which returns
a pair of |int|s, where the first is |nMaxIn| and the second is |pMaxIn|.

In the |analyze| step of the analysis, correlators can be filled with an applied |Correlators| projection. The projection is applied as usual e.g.~by
|const Correlators& c = apply<Correlators>(event, "ProjectionName");|, and a booked $m,n$-correlator is filled as |corrPtr->fill(c);|.

In the |finalize| step of the analysis, correlators can be cast into cumulants or flow coefficients. If an analysis implements e.g.~experimental data
on integrated $c_2\{2\}$ and $v_2\{2\}$, the methods |cnTwoInt(Scatter2DPtr h, ECorrPtr e2)| and |vnTwoInt(Scatter2DPtr h, e2)| maps the correlator |e2| to
scatters pointed to by |h|.

\subsection{Deep-inelastic $ep$ scattering and photoproduction}\label{sec:hera}

Although \rivet traces its conceptual origins directly back to
\hztool~\cite{Bromley:1995np,Waugh:2006ip}, a \fortran package
developed by the H1 and ZEUS collaborations at the HERA $ep$ collider
to facilitate the comparison of their measurements to MC predictions
and each other, rather few deep inelastic scattering (DIS) or
photoproduction measurements have been implemented in \rivet to
date. This is partly because of the existing and extensive library of
such analyses in \hztool.  Such measurements contain important and unique
information, principally on high-energy QCD dynamics and hadronic
structure, which remains relevant to current and future analyses.
The need to preserve access for use with
modern event generators, and to exploit the ongoing benefits of new
development in \rivet --- several of which are informed by lessons
learned from \hztool --- has grown over the years, and has been
further stimulated by the active community work toward a future
electron--ion collider. As a consequence efforts have been made to
interface the old \hztool Fortran routines to \rivet, and a proper
plug-in library is all but released~\cite{RivetHZtool}. In parallel to
this the latest version of \rivet contains a few more HERA analyses,
but more importantly it provides a series of new projections to extract
common DIS and photoproduction event properties, greatly facilitating
the addition of new (or indeed old) DIS and photoproduction analyses
in future.
The currently available HERA routines are given in Table~\ref{table:heraanalyses}.

\subsubsection{Kinematic definitions}

In following the \rivet philosophy of defining observable in terms of
final state particles, to avoid model- and generator-dependence, a
number of physics issues arise which were in general noted, but not
solved, in \hztool. In particular, it is not always clear from the H1
and ZEUS analyses how to do the following, without examining the
(unphysical and generator-dependent) parton-level event history.

\paragraph{Identifying the scattered lepton}
In most DIS events, there is a single obvious scattered lepton candidate. However, other leptons may be present in the event
and \rivet (and in principle physics!) require a way resolving any ambiguity based upon observable information --- essentially the kinematics.
Unfortunately most HERA publications do not provide this information (and indeed were often corrected to the MC-dependent electron vertex).
The |DISLepton| projection therefore provides a few of pragmatic recipes to identify the scattered lepton, with options to select
the highest energy lepton (default) or by rapidity or transverse energy. The electron may also be required to be prompt (default).

The scattered neutrino kinematics in charged current events should be determined from the missing transverse energy. However, no such analyses are
currently implemented.

\paragraph{Treating electroweak corrections to the Born-level process}
Many HERA analyses were corrected to the ``Born'' level, again leaving some ambiguity about how radiated photons should be treated, when they
are present. Of course, events may be run at fixed order with QED radiation turned off, and, while model-dependent in principle, this is
most likely the closest approximation to what was done in the original measurement. To allow the study of such effects, the
|DISLepton| projection will, if requested, return the kinematics of the electron including in the lepton four-momentum all photons within some
cone (thus recovering final-state QED radiation to some approximation), and excluding from the energy of the beam electron all photons in some
cone (this accounting at some level initial state radiation). A hadronic isolation criterion may also be applied.

The |DISKinematics| projection then calculates the usual DIS variables, such as Bj\"orken $x$, $y$ and $Q^2$, from the identified scattered electron
and beam energy. The |DISFinalState| returns the final state particles excluding the scattered lepton, optionally boosted into the hadronic centre-of-mass
frame, the Breit frame, or left in the laboratory frame.

\paragraph{Identifying the photon kinematics in photoproduction}
Photoproduction is viewed as a special, low-$Q^2$ case of DIS. In most analyses, a veto is applied on the scattered electron entering
the detector acceptance, typically corresponding to an upper cut on $Q^2$ of 1--4~GeV$^2$. The |DISKinematics| projection may thus be used
to obtain the energy and virtuality of the interacting photon.

\paragraph{Defining diffraction and its kinematics}
For diffractive analyses with a tagged forward proton, the issues are similar in principle to those associated with
identifying the scattered lepton, but in practice the highest rapidity proton is always identified by the |DISDiffHadron| projection.
In other diffractive analyses, the diffractive final state is identified by the presence of a rapidity gap amongst the hadrons.
A |DISRapidityGap| projection exists to facilitate this.

\begin{table}
  \begin{adjustbox}{width=\textwidth,center}
  \begin{tabular}{llllr}
	\toprule
	Analysis name & System & Experiment & Measurement features & Reference \\
	\midrule
	\verb:H1_1994_S2919893: & $e^\pm p$ DIS & H1 & Energy flow and charged particle spectra  & \cite{Abt:1994ye} \\
	\verb:H1_1995_S3167097: & $e^\pm p$ DIS & H1 & Transverse energy and forward jet, low-$x$ & \cite{Aid:1995we} \\
	\verb:H1_2000_S4129130: & $e^\pm p$ DIS & H1 & Energy flow & \cite{Adloff:1999ws} \\
	\verb:H1_2007_I746380:  & $e^\pm p$ DIS \& \\
                                & $ep \rightarrow \gamma p$ & H1 & Diffractive production of dijets  & \cite{Aktas:2007hn} \\
	\verb:H1_2015_I1343110: & $e^\pm p$ DIS & H1 & Diffractive dijets & \cite{Andreev:2015cwa} \\
	\verb:HERA_2015_I1353667: & $e^\pm p$ DIS & H1\& ZEUS & Combined H1/ZEUS $D^*$ production  & \cite{H1:2015dma} \\
	\verb:ZEUS_2001_S4815815: & $ep \rightarrow \gamma p$ & ZEUS & Dijets  & \cite{Chekanov:2001bw} \\
	\verb:ZEUS_2008_I763404:  & $ep \rightarrow \gamma p$ & ZEUS & Diffractive photoproduction of dijets & \cite{Chekanov:2007rh} \\
	\verb:ZEUS_2012_I1116258: & $ep \rightarrow \gamma p$ & ZEUS & Inclusive jet cross sections  & \cite{Abramowicz:2012jz} \\
	\bottomrule
  \end{tabular}
  \end{adjustbox}
  \caption{\label{table:heraanalyses}
All \rivet analyses routines available in a reasonable state of usability (though in general not formally validated).
An up-to-date list can be found at \texttt{https://rivet.hepforge.org}.
}
\end{table}

\subsection{Detector emulation}
\label{sec:detector}

\rivet was initially developed to encode \emph{unfolded} analyses, i.e.~those
for which the biases and inefficiencies introduced by detector material
interactions and imperfect reconstruction algorithms have been corrected, making
the experiment's published observables the best possible estimate of what
happened 
at the fundamental interaction point, independent of any particular detector.

It remains our firm belief that unfolded measurements --- while requiring a
significant investment of time and effort to understand and invert the detector
biases, and to minimise model-dependence in the derivation of such corrections
--- are the gold standard form in which to publish collider physics
measurements. This is particularly the case when the fiducial analysis
phase-space (i.e.~the allowed kinematic configurations at truth-particle level)
has been carefully designed to minimise extrapolation beyond what the detector
could actually (if imperfectly) observe.

Not all collider physics analyses are appropriate for detector-unfolding,
however. For example, stable unfolding relies on probabilistic algorithms to
determine the probabilities of event migration between truth-particle and
reconstruction-level observable bins, and hence the MC populations used to
derive these migration probabilities must be large enough to achieve statistical
convergence. Some analysis phase-spaces, in particular BSM particle searches on
the tails of distributions or on the edges of allowed kinematics, may be
prohibitively difficult to simulate in the required numbers. Even if the MC
samples can be made sufficiently large, the propagation of small numbers of
observed events through the unfolding machinery can be fatally unstable, and
also the low number of events present in the data means the ability to validate
a large MC sample may be limited, unless appropriate control regions can be
defined. Finally, the culture of BSM searches has historically been that speed
is of the essence in the competition between experiments. Therefore, as
unfolding --- with its additional complexity and need for extensive
cross-checking --- does not intrinsically add exclusion power in e.g.~the
studies of simplified BSM models that LHC experiments use ubiquitously as
phenomenological demonstrations of analysis impact, it has typically been
neglected from search analyses.  While this culture is necessarily changing in
the high-statistics limit of LHC running, where an extra 6 months of data-taking
does not automatically revolutionise the previous measurements, and in the
realisation that simplified models are not always good proxies for full
UV-complete BSM models~\cite{Athron:2018vxy,Amrith:2018yfb}, it remains the case
that with a few exceptions~\cite{Aaboud:2017buf,Aaboud:2019jcc}, unfolding is
currently rarely part of the vocabulary of collider BSM direct-search analyses.

It is in response to these factors that machinery for detector emulation has
been added to \rivet --- to permit the important class of reconstruction-level
search analyses to be preserved for re-interpretation, albeit through an
approximate detector model. The detailed implementation of this is reviewed in
Ref.~\cite{Buckley:2019stt}, along with comparisons to other
initiatives~\cite{Drees:2013wra,Conte:2012fm,Balazs:2017moi} with similar intentions,
but here we give a brief flavour of the features.

The key decision in implementing detector modelling was to use a ``smearing +
efficiency'' approach rather than to attempt to model detector geometries,
particle--material interactions, and thousands of lines of private
reconstruction software within the \rivet package. This long-established
approach distinguishes the \rivet
detector modelling from that of the \textsc{Delphes} fast-simulation
code~\cite{deFavereau:2013fsa}. Specifically, we have chosen to implement detector effects as ``wrapper''
|SmearedParticles|, |SmearedJets|, and |SmearedMET| projections around the
standard particle-level |ParticleFinder|{,} |JetFinder|, and |MissingMomentum|
classes. These wrappers perform the dual tasks of modelling reconstruction
efficiency losses, such that the wrapper may return a sampled subset of the
particles (or jets) found by the contained truth-particle projection, and
(except for the MET one) of ``smearing'' the 4-momenta of the truth-level
objects to represent inaccuracies in kinematic reconstruction. Both the
efficiency and smearing decisions are made using user-supplied functors
(i.e.~named functions, lambda functions, or stateful function objects) of the
physics objects, respectively returning either a |bool| for efficiency filtering
or a new |Particle|/|Jet| for smearing, with the sampled loss rate and smearing
distributions dependent on the original object properties, most commonly their
\abseta and \pt. The advantage of this method, in addition to simplicity, is
that it is fully customisable to measured efficiency and smearing effects in the
specific phase-space of each analysis, and can be embedded directly in the
analysis code, rather than relying on the universal correctness of a monolithic
third-party detector simulation.

In addition to this machinery, a large number of standard efficiency and
smearing parametrisations for ATLAS and CMS have been implemented, based on a
mix of \textsc{Delphes} configurations and experiment reconstruction performance
papers~\cite{ATLAS:2015uwa,ATL-PHYS-PUB-2015-041,ATLAS-CONF-2016-024,Aaboud:2019ynx,ATL-PHYS-PUB-2015-037,Aad:2016jkr,Aad:2014rga,ATL-PHYS-PUB-2015-045,Aaboud:2016yuq,ATL-PHYS-PUB-2016-014,Aad:2012re,Khachatryan:2014gga,CMS-PAS-JME-17-001}.
These in turn are based on generic helper functions such as Gaussian \pt or mass
smearers, $b$-tag efficiency/fake samplers, etc., which also act as a useful
foundation on which users can build their own detector parametrisations. As with
all \rivet tools, the implementation emphasises well-behaved default settings,
and physics content over language noise.


\subsection{BSM search analysis features: cut-flow monitoring}
\label{sec:bsm}

The object-filtering metafunctions and detector emulations described above
constitute the main features that allow \rivet to now support the majority of
LHC BSM search analyses. For example a search analysis can find leptons with
e.g.~|FinalState(Cuts::abspid == PID::MUON && Cuts::pT > 50*GeV && Cuts::abseta < 2.7)|,
or |DressedLeptons(PromptFinalState(Cuts::abspid == PID::ELECTRON))|,
then wrap them into a detector-biased form with 
a call to %
e.g.~|SmearedParticles(| |elecs,| |ATLAS_RUN2_ELECTRON_EFF_TIGHT,| |ATLAS_RUN2_ELECTRON_SMEAR)|,
which is then declared and applied like a normal
particle-level projection. Jets found and smeared similarly can be isolated from
the leptons using a few calls to functions like |discardIfAnyDeltaRLess(elecs, jets, 0.4)|.
This makes for an expressive and powerful reconstruction-level
analysis emulator, comparable to other tools on the market.

\rivet provides one more feature specifically targeted at BSM search
implementation: a dedicated cut-flow monitoring tool. All analyses apply chains
of event-selection cuts, but these are particularly crucial for re-implementers
of search analyses because the cut-flow --- i.e.~the sequence of numbers or
fractions of signal events passing each selection cut --- is often the only
published validation metric comparable in detail to the differential histograms
of measurement analyses. Coding a cut-flow monitor by hand, e.g.~via a
histogram, is easy enough, but rather repetitive: one often has to write the cut
logic once for the histogram fill, and then again for the actual event veto or
signal-region iteration. The problem becomes particularly acute when, as is
often the case, the analysis contains many parallel search regions, all of which
have their own cut-flow.  On top of all this, one needs to then code a print-out
of the cut-flow stages, including raw event counts --- possibly normalised to
cross-section and luminosity, or to a fixed reference --- as well as step-wise
and cumulative efficiency fractions.

\rivet's |Cutflow| object exists to solve these problems. It acts like a
fillable histogram, augmented with extra methods to attach human-readable labels
to each cut stage, to track the current cut stage rather than make the user do
so manually via a histogram's fill variable, to be able to fill multiple cut
stages at once, and to return the result of a fill such that it can
simultaneously update the cutflow and a signal-region counter, e.g.\
|signalregion[i]->fill(_cutflow->fillnext({pT1 > 50*GeV, aplanarity < 0.1}))|.
Passing to a |stringstream| produces a nicely aligned text
representation which can be printed with e.g.~|MSG_INFO(_mycutflow)| in the
analysis's |finalize()| stage, with the raw-count, step-wise efficiency, and
cumulative efficiency measures all reported. |Cutflow::scale()| and
|normalize()| methods are provided, the latter with a optional flag to determine
which cut stage the normalization should be matched to. In addition, a map-like
|Cutflows| wrapper is provided, for containing many named |Cutflow| objects, and
potentially filling all of them with a single |fill()| call to the container; it
is also able to write out its contained cut-flows in the standard form. These
small but significant features make debugging and validating a search analysis a
more pleasant experience.

\subsection{Analysis options}
\label{sec:analysis-options}

From the beginning it was assumed that an analysis in \rivet is
completely specified by the corresponding |Analysis| class and could
only be run in one way. However, especially with the introduction of
heavy ion analyses, it became clear that some analysis has to be
treated differently depending on which event generator it is used
for. In particular for the centrality concept used in heavy ion
analyses, there are different ways of handling them (see Section~\ref{sec:hion}).
Similar issues arise for analyses based on a single paper, but in
which cross-sections are measured for e.g.~muons and electrons
separately and also combined, or for different definitions of heavy-flavour-tagged
jets. In such cases it is efficient to be able to specify different running
modes for a single analysis.


For this reason \rivet now includes an option machinery. For any
analysis name it is possible to specify one or more options that can
take on different values. This is encoded by supplying suffixes to the
analysis names on the form\\
|  rivet -a AnalysisName:Opt1=val1:Opt2=val2|\\
In the analysis class it is then possible to retrieve the specified
value of a particular option with the function %
|string Analysis::getOption(string)|.

It is possible to specify the options and values that are allowed
for a given analysis in the |.info| file, but it is also possible
communicate other options to an analysis.

Note that it is possible to use several versions of the same analysis
in the same \rivet run as long as the versions have different options.

Handlers are provided in the \python interface to extract or strip the
options from histogram paths.

\subsection{Dependency evolution}
\label{sec:other}

In addition to explicit \rivet framework developments, version~3 supports new
versions of dependencies, both physics libraries and the Python runtime.

\subsubsection*{HepMC3}
\label{sec:hepmc3}

The default configuration of \rivet currently assumes that HepMC
version 2 is used for the event files to be read by \rivet. In the
future this will be changed to compiling \rivet with HepMC3. Already
now it is possible to try out HepMC3 with rivet by providing the flag
|--with-hepmc3=/path/to/installation| to the |configure| script.

Note that when compiled with HepMC3, Rivet can still read HepMC2
files. In fact \rivet will then automatically detect the format the
given event file.

\subsubsection*{\python~3}
\label{sec:python-3}

While the \rivet core library and the analysis implementations are written
in \cxx, there is a full featured Python wrapper around them built using the
Cython system. In fact, the |rivet| executable is written in Python and uses
this interface.

Cython is not a requirement to install the \rivet Python modules if
building from a release tarball, but is necessary when building from the git
repository directly.

Recent \rivet versions are compatible both with Python~2.7 and Python~3.
If multiple Python versions are available in a system, you can choose which one
to use for the \rivet installation by prefixing the configure command with e.g.\
|PYTHON=/usr/bin/python2 ./configure ...|.
It is also possible to install modules for both versions in the same
installation location by running the appropriate
|PYTHON=<python> ./configure ...; make install| twice.

\section{User guide}
\label{sec:userguide}
Here we provide a short user guide to help with getting \rivet~3 up and running
both standard built-in analyses, and your first analysis routines. Full
information, including full code API documentation, can be found on the \rivet
website at \url{https://rivet.hepforge.org/}, and the \yoda one at
\url{https://yoda.hepforge.org/}.

\subsection{Installation}
Getting started \rivet is most easily done using the \textsc{Docker} images
installed via |docker pull hepstore/rivet|, then entering an interactive
environment with \rivet available %
with e.g.~|docker run -it hepstore/rivet|. Many useful variants on these
commands can be made, %
for example using an image with an MC generator also included, such as
|hepstore/rivet-pythia|, %
or using the |-v| flag to |docker run| to mount host-system directories inside
the image so external event files can be easily read in, and histogram files
written back out. Alternatively, a native installation can be made easily on
most *nix systems by downloading a ``bootstrap script'' which will build \rivet
and all its dependencies --- full instructions for this are provided on the
website. After installation and set-up of the \rivet environment, it can be
accessed at the command-line using the command |rivet|: for example, try
getting the list of supported commands and options using |rivet --help|.

\subsection{Analysis metadata inspection}
The first point of call in using \rivet is finding which analyses are of
interest. There are over 900 analyses in the current release, as you can verify
by running |rivet --list-analyses| and counting the number of resulting lines,
and so being able to find the ones you want requires some searching. A full list
of standard analysis routines, with information about the collider, energy,
luminosity, date, process type, an abstract-like description, bibliographic
data, and a syntax-highlighted copy of the analysis code, can be found on the
\rivet website. The information used to build these web pages is also accessible
from the command-line, with for example |rivet --list-analyses ATLAS_2018| being
usable to print a one-line description of each routine whose name contains the
pattern ``ATLAS\_2018'' (corresponding to ATLAS experiment papers published in
2018), and |rivet --show-analyses ATLAS_2018| printing out the fully detailed
metadata entries for each analysis matching the pattern. You will notice that
the majority of analyses have a standard name of this kind:
$\langle\mathit{expt}\rangle\text{\_}\langle\mathit{year}\rangle\text{\_}\langle\text{I}\mathit{nnnnnnnn}\rangle$,
where the last portion is an ID code corresponding to the analyses key in the
Inspire publication database~\cite{Inspire}.

\subsection{First \rivet runs}
Now to run analyses on simulated events, using pre-written analyses. Let's say
we want to analyse some top-quark pair ($t\bar{t}$) events at
$\sqrt{s} = 13~\TeV$, using routines that both compare to real data, and
generically characterise the jet and partonic top distributions in the
events. Using the metadata search system on the command-line we find the
|CMS_2018_I1662081| data analysis, and also the |MC_JETS| and |MC_TTBAR|
routines which have no experimental analysis counterpart. These analyses are
located by the \rivet analysis loader system, which by default looks in the
\rivet library install path under |$prefix/lib/Rivet/|. 
If the |$RIVET_ANALYSIS_PATH| 
environment variable is set, or search paths are specified via the \rivet
library API, these are used by preference with fallback to the default unless the
path variable ends with a double-colon, |::|.

If you have a set of 10k--1M $t\bar{t}$ events in HepMC format, then running is
trivial --- just tell |rivet| which event file(s) to read from, and which
analyses to run, e.g.~|rivet -a CMS_2018_I1662081 -a MC_JETS,MC_TTBAR events.hepmc|. %
Analyses can be specified both by multiple instances of the |-a| option flag, or
by comma-separating analysis names in the arguments to a single |-a|, as
shown. The event file may be either uncompressed, or gzipped, but must be
supported by the HepMC library.

More commonly, especially for large MC event samples, we generate the
parton-showered events ``on the fly'', and pass them directly to \rivet.  This
can be done most efficiently by using the \rivet \cxx library API to hand HepMC
objects in memory between the event generator and \rivet, and so requires either
built-in \rivet support in the generator (as for \sherpa~\cite{Gleisberg:2008ta}
and \herwig~\cite{Bellm:2015jjp}), or for
the user to write a \cxx program that uses both libraries (as is the case with
\pythia\,8). A slower, but completely generator-agnostic, way is to write out a
temporary event file and read it in to \rivet: for this, Unix systems have a
very useful feature in the form of a ``FIFO file'' --- a file-like object for
inter-process communication. To run \rivet this way, first make a FIFO with
e.g.~|mkfifo myfifo.hepmc|, then run the generator \emph{in the background} (or
in a separate terminal on the same system) with instructions to write
HepMC-format events out to the FIFO: %
|some-generator --some-config=ttbar --out=myfifo.hepmc|.  %
Finally, run |rivet| as before: the generator writing and \rivet reading will
control each other such that events are passed between them and the ``file''
never gets any bigger than a few tens of kilobytes.

\rivet will happily chug through the provided events, updating an event counter
on the terminal and periodically writing out a |.yoda| file containing output
histograms and counters from the |analyze()| and |finalize()| stages of the
\rivet analyses' processing. Using the tools described in the following section,
you can inspect and plot these intermediate files should you wish. If you find
that you have acquired sufficient statistics, and don't need the rest of the
generator file or run, you can perform the usual |Ctrl-C|
intervention 
to kill the |rivet| process, which will exit gracefully after |finalize|ing the
remainder of |analyze|d events.

\subsection{Plotting and manipulating results}
The usual next step is to plot the results. The final |.yoda| file written by
the |rivet| run (named |Rivet.yoda| by default) is the principle input to this
plotting, optionally along with equivalent files from other MC runs.

If multiple MC runs --- either for separate MC processes or to split a large
single homogeneous-process run into smaller chunks --- need to be combined into
a single |.yoda| file for this purpose, the |rivet-merge| script can be used to
read in these independent contributions and re-run the analyses' |finalize()|
methods to give a final, statistically exact combined |.yoda| file, as described
in Section~\ref{sec:reentrant-finalize}. Cross-section and number-of-events
scalings will be automatically calculated from information stored in the input
files. Should any manual scaling be needed in addition, the |yodascale| script
or a custom manipulation using the YODA Python API are also possible.

The usual approach to plotting is to run the |rivet-mkhtml| script. This is a
wrapper around the lower-level |rivet-cmphistos| and |make-plots| scripts, which
respectively group sets of histograms by analysis, and render them to PDF
format, with additional generation of HTML code and thumbnail images so the
output can be conveniently viewed via a Web browser. Reference data will be
automatically loaded from the same location as the compiled analysis library (or
more generally from the |$RIVET_DATA_PATH| path list). 


\subsection{Basic analysis writing}
The writing of analyses is as potentially multi-faceted as writing any computer
program, and hence cannot be covered here in comprehensive detail. The best way
to learn, as ever, is by doing and by consulting existing analysis routines with
similar ambitions to your own. But compared to a completely general program,
\rivet routines are constrained by the three-step |init|/|analyze|/|finalize|
event-processing structure, and by their necessary inheritance from the
|Rivet::Analysis| type: here we will survey the main features of each step.

\subsubsection*{Raw materials}
Our first step is to generate the file templates into which we will insert
analysis data and logic. The starting point should always be the
|rivet-mkanalysis| script, run like |rivet-mkanalysis EXPT_2019_I123456| where
the analysis name follows the three-part structure described
earlier. Particularly critical is the third part, encoding the Inspire database
key with which the script can automatically extract reference data and
publication metadata.

Running the script in this way generally results in four template files: a |.cc|
file containing a generic template for your analysis code, with ``boilerplate''
features pre-filled; a |.info| metadata file in YAML format, used to generate
documentation and constrain applicability; a |.plot| file used to apply plot
styling directives to sets of output histograms; and a |.yoda| reference data
file in the YODA format, downloaded if possible from the HepData database. The
only essential file for private running is (of course) the |.cc| in which the
analysis logic will be written, but any analysis submitted to \rivet for
official inclusion in the repository must also complete the other files. In
particular it is critically important that the |.yoda| file be compatible with
HepData's YODA output, so updates can be synchronised with subsequent \rivet
releases.

\subsubsection*{Projections}
Projections are the engines of \rivet: they are calculators of observables,
encapsulating various nuances in the definitions and efficiency insights for
calculation, as well as benefiting from the automatic caching of their results.
The most important projections are those which inherit from the |ParticleFinder|
interface: these include the |FinalState| projection which extracts and returns
subsets of stable final-state particles; its specialised children like
|PromptFinalState| which excludes final-state particles from hadron decays, and
|VisibleFinalState| which only returns particles that would interact with a
detector; composite final-state particle finders like |DressedLeptons| which
sums prompt photons in cones around charged leptons; decayed particle finders
|UnstableParticles| and |TauFinder|; and pseudoparticle finders like |WFinder|
and |ZFinder| which reconstruct leptonic composites at EW scale by
experiment-like fiducial definitions. Other important projections are
|FastJets|, the main implementation of the |JetFinder| interface, and
|MissingMomentum| for a particle-level definition of missing $E_T$. Using these
projections affords the analysis author quick and robust definitions of physics
objects and quantities from which the rest of the analysis logic can be applied.

The caching machinery around projections means that they must be used in a
slightly non-intuitive way: they are \emph{declared} in the |init()| method of
an analysis, and then retrieved and \emph{applied} in the |analyze()| step. The
declaration involves first constructing and configuring a local object in the
|init()| method, e.g.~|FinalState fs(Cuts::pT > 10*GeV);| and then assigning a
string name to it, e.g.~|declare(fs, "MyFS");|. The string name must be unique
within this analysis, but different analyses are free to use the same
names. Once declared, the projection object has been cloned into the \rivet
core, and the local copy will be automatically deleted once the |init()| method
closes. Then in |analyze()|, the projection's computation is performed by
referencing the registered name, %
e.g.~|FinalState& pf = apply<FinalState>(event, "MyFS");|. In fact it is common
to bypass the projection itself in the application, going straight to the result
of one of its methods, %
e.g.~|const Particles ps = apply<FinalState>.particles();|.

\subsubsection*{Histograms and counters}
Statistics objects in \rivet must, like projections, be managed by the system
core: this is to enable the automatic handling of event-weight vectors,
including details such as fractional fills and counter-event groups
(cf.~Section~\ref{sec:weights}), as well as run-merging and re-entrant calls to
the |finalize()| function. For efficiency, convenience and flexibility in how
they are handled by user code, references to \yoda histograms, profile
histograms, and weight counters are stored as |Rivet::Histo1DPtr|,
|Rivet::Profile1DPtr|, |Rivet::CounterPtr|, etc.\ member variables within the
analysis class. The actual histogram configuration is performed on these using a
set of overloaded |Analysis::book(hptr, ...)| methods, where |hptr| is any of
these |Rivet::*Ptr| objects. For binned histograms, the remaining arguments can
be a \cernroot-style (\textit{hname}, $N_\text{bins}$, $x_\text{min}$,
$x_\text{max}$) tuple, a (\textit{hname}, $[x_\text{edges}]$) pair, or a single
name string corresponding to the reference data histogram whose binning should
be used by its \rivet equivalent. This latter form also has an integer-triplet
shorthand, expanding to the |daa-xbb-ycc| dataset/axes format output by HepData.
Counters and |Scatter*D| objects, which do not have a variable binning, have
simpler |book()| method overloads. Note that analyses which run in several
modes, e.g.~making the same kinds of observable histograms for event runs at two
different $\sqrt{s}$ energies, so not need different histogram pointer variables
for each mode --- simply pass different additional arguments to the |book()|
methods depending on the context of, for example, a call to
|Analysis::sqrtS()/GeV|.

Within the analysis, the |Ptr| objects are used in a straightforward fashion,
most notably calls like e.g.~|_h_myhist->fill(x)|. Users of \rivet~v2 will note
that the event weight is no longer part of this function signature. In fact,
attempts to call |Event::weight()| will now be spurned, returning |1.0| and a
warning message. This is because in \rivet~v3, there is no single event weight
and it is the job of the \rivet \emph{system} rather than the user to handle and
combine weight vectors correctly. Weighted fills are still allowed, but for
weights than the event ones. Behind the scenes, the |Ptr| objects are
multiplexed on to arrays of fundamental \yoda objects, with \rivet secretly
looping over weight arrays or running methods once for each weight, but users
can and should act happily oblivious to these sleights of hand.

The biggest such trick occurs between the |analyze()| and |finalize()| methods,
when the active histograms (or histogram \emph{sets}) are persisted as ``raw''
versions, to allow for pre-finalize run combination. Within finalize, each
weight stream is treated independently, but again users need not worry about
these details. The most common finalising operations are calls to |normalize()|
or to |scale()|, which respectively fix the area of a histogram to a fixed
number (perhaps proportional to the process cross-section obtained via
e.g.~|crossSection()/femtobarn|) or scale it by a factor. In analyses where not
every event results in a histogram fill, the latter approach is usually what is
wanted if measuring absolute cross-sections, e.g. %
|scale(_h_myhist, crossSection()/picobarn/sumW())|: this division of the
histogram's accumulated sum of cut-passing event weights by the all-event
process cross-section |sumW()| encodes an acceptance factor for each
weight-stream into the final histogram normalization. |Counter| objects may also
be useful for this acceptance tracking. For convenience, the |normalize()| and
|scale()| methods also accept histogram containers, either explicit or via an
initialisation list %
cf.~|normalize({_h_pt, _h_eta, _h_mass})|. Both the raw and finalized statistics
objects are automatically written out, both periodically and at the end of the
run, exactly as for ``official'' \rivet routines.

\subsubsection*{Analysis logic and tools}
Having established the analysis class structures, and the core machinery of
declared projections and booked histograms, the ``only'' remaining part is the
logic of your specific analysis as acted out with these participants. Here there
is relatively little to say: the principle logic and control flow tools are
simply the syntax of procedural/object-oriented \cxx: |for| and |while| loops,
|if ... else| statements, boolean expressions, ternary |x ? y : z| syntax, etc.

Any feature of the STL can also be used, with standard types like |std::vector|
and |std::map| already imported into the |Rivet| namespace and more clearly
referred to simply as |vector| and |map|. For convenience, vectors of some
standard types are given convenience aliases, e.g.~|Particles|, |Jets|,
|doubles|, |strings|, and |ints|. The first two in this list are \rivet-specific
types which it is worth becoming familiar with in some detail, as they support
not just kinematic operations like |pT()|, |eta()|, |absrap()|, etc.\ (as
described earlier in this document), but also ways to interrogate their
composite nature, decay ancestry, connection to |HepMC::GenParticlePtr| objects,
etc. |Jet| objects additionally allow for $b$- and $c$-hadron truth-flavour
labelling using a robust and now-standard ghost-association definition.

Note that all unitful kinematic quantities such as |ParticleBase::pT()|, |mT()|,
etc.\ (as well as the return value of |Analysis::crossSection()|) are returned as
|double|s, but should be treated in this form as having an undefined ``default
\rivet unit'': they are not safe to be passed to a \yoda histogram without an
explicit unit declaration. The definition used for this is that multiplying by a
unit constant converts the numerical value from that unit to the \rivet default
unit, and dividing will convert back to a number representing how many of the
dividing unit the \rivet internal value corresponded to. Hence the RHS term in
|Cut::pT > 10*GeV| converts from ten \GeV units to \rivet's internal scheme, and
|hist->fill(p.pT()/GeV)| is a conversion to the unitless number to be plotted on
a \GeV axis.

Several other additional features have already been described, such as the
filtering |select()|, |reject()|, |iselect()|, etc.\ functions, and the many
functors to be used with them. Many examples of best-practice usage may be found
in the code and in dedicated tutorials available from the \rivet Web page.

\subsection{Building and running analysis codes}
Having written the \cxx analysis code, it must be built into a compiled
``plugin'' library that can be loaded by the \rivet system. This is done at
runtime without needing the core \rivet framework to be recompiled, using the C
|dlopen()| dynamic loader. Since the core library and plugin must be compatible
at compiled binary code level, the \cxx compiler used for plugin building must
see exactly the same headers from \rivet dependency packages like HepMC and
FastJet: this leads to a complex compiler command line that could easily be a
source of mysterious technical crashes or strange behaviour, and hence a
convenience wrapper script, |rivet-build|, is provided to encode all the
necessary compiler incantations. It is run on any number of analysis source
files like %
|rivet-build MYANA1.cc MYANA2.cc ... -lextralib -extra_option| %
and produces a |dlopen|able shared library named |RivetAnalysis.so|.

Obviously it would be awkward if all analysis plugin libraries had to have the
same filename, and so a custom output name can be given as an option first
argument in the form |Rivet*.so|. When running \rivet, the library will search
the |$RIVET_ANALYSIS_PATH| 
variable and installation prefix fallback (as previously described) for |.so|
libraries matching this pattern, from which all the contained analyses will
register themselves with the \rivet core. As a convenient shorthand for the path
variable setting, the |rivet| script (and related tools like |rivet-mkhtml| and
|rivet-merge|) can take an optional |--pwd| flag, equivalent to prepending
|$PWD| 
to the analysis and data search paths.

\subsection{Contributing analysis routines}
We encourage all users to contribute validated analyses to the official \rivet
collection: this helps the whole particle physics community, and your efforts
will be accredited through public analysis authorship declarations. After
submission, the \rivet core authors will be responsible for maintaining the
analysis code compatibility with any future changes in the core API.

In addition to the |.cc| code file, the metadata |.info| file, plot styling
|.plot| file, and if appropriate |.yoda| reference data files must be provided,
along with information (ideally including plots from |rivet-mkhtml|)
illustrating the pre-submission validation procedure performed by the analysis
author. The info file must include a |ReleaseTests| entry indicating how a short
analysis-validation behavioural regression run should be performed, using the
example 1000-event HepMC event files located at
\url{http://rivetval.web.cern.ch/rivetval/}. If
no existing event file is suitable for the analysis, a new |.hepmc.gz| analysis file
should be supplied along with the analysis code and metadata upload.

In the past, contribution of analyses and this supporting validation information
has been done through a mix of email submissions to the \rivet developer mailing
list and (for official experiment representatives) upload to a special
``contrib'' area on the \rivet website. Since version 3.0.2, a more formal
procedure is in operation, whereby new analyses are to be contributed via merge
requests on the
\url{https://www.gitlab.com/hepcedar/rivet} code
repository. Validation plots and similar information, and new HepMC event
samples if appropriate, should be contributed to the \rivet core team separately
from the repository merge request.

Our thanks in advance for contributing to this important community analysis
preservation resource!

\section{Conclusions and future plans}

Over the last decade, \rivet has become established in the ecosystem of particle
physics analysis preservation, primarily for, but not limited to, the LHC
experiments. Its position in this world is an intermediate one, using more
complex and complete particle-level final states than in partonic matrix-element
analysis tools, while less fully detailed (and computationally expensive) than
forensic preservations of experiment simulation and analysis frameworks. This
mixture of detail and efficiency has led to many uses in studies from MC
generator model and tune optimisation, to limit-setting studies on explicit and
effective Beyond Standard Model theories.

In this review we have noted how the core set of ``\rivet{}able'' analyses,
formalised as fiducial phase-space definitions, have become part of standard
operating procedure for detector-corrected measurements, while the remit of
\rivet has expanded to include more complex multi-pass observables as used in
heavy ion physics, and approximations of reconstruction-level quantities,
particularly new-physics searches. \rivet has also evolved to take advantage of
new developments in the precision and control over calculational systematics
from Monte Carlo event generators, in the form of an unprecedentedly transparent
handling system for event weights and counter-event groups. The full realisation
of these features, and extensions to yet more areas of particle physics such as
astroparticle and neutrino physics, is a challenge for the next decade of \rivet
and of particle-physics data- and analysis preservation.

\section*{Acknowledgements}
We thank all contributors of analysis codes, bug reports, and code patches and
improvement suggestions for their efforts to make \rivet the best possible tool
for analysis preservation and interpretation. Our thanks also to the MC
generator author community for their support of the \rivet project, and all
those who hosted the meetings and workshops that informed and accelerated
its development.

\paragraph{Author contributions}
Project leadership, and evolution of fiducial analysis concepts: AB, JMB. %
Core machinery, cuts system, and projection design and implementation: AB, LC, LL, DG, CG, HS, FS. %
Multi-weight, re-entrant finalize, and histogramming system: AB, LC, DG, CG, CP, LL. %
Design and development of heavy ion analysis features: CB, CHC, AB, JFG-O, PK, JK, LL. %
Refinement of $ep$ physics tools and definitions: AB, JMB, LL. %
Detector emulation system and search tools cf. filtering functors: AB. %
Validation and low-$Q$ / hadronic decay physics coverage: PR %

\paragraph{Funding information}
This work was supported in part by the European Union as part of the Marie
Sklodowska-Curie Innovative Training Network MCnetITN3 (grant agreement
no. 722104).
AB thanks The Royal Society for University Research Fellowship grant
UF160548. AB and CP thank the University of Glasgow for postdoctoral funding
through the Leadership Fellow scheme. LL was supported in part by Swedish
Research Council, contract number 2016-03291. FS was supported by the German
Research Foundation under grant No.\ SI 2009/1-1.  Many thanks to Xavier
Janssen, Markus Seidel, Alex Grecu, and Antonin Maire for (in addition to
members of the author list) acting as LHC experiment
contacts. 

\begin{appendix}

\section{Handling groups of completely correlated events}
\label{sec:handl-groups-compl}

The problem addressed here is how to process histogram fills in an event group
with fully correlated events, in a way such that the it is treated as one event
fill and still have the correct error propagation in terms of
the 
\sosw. It is also essential to make sure that large cancellations remain
cancelled across bin edges. The solution is to introduce \textit{bin smearing}
and the concept of a \textit{fractional fill}.

As of \yoda version~1.6, the histograms have an extra argument to their
fill functions, encoding the concept of a fractional fill. This means
that instead of having one fill with weight $w$, we divide it up in
$n$ fills with weights $f_iw$ where $\sum f_i=1$. For the single fill
we will add $w$ to the \sow and $w^2$ to the \sosw. The \sow is no
problem, we simply add up the fractional fills: $\sum f_iw=w$, but the
naive approach of doing $n$ fractional fills would give a contribution
$\sum (f_iw)^2\ne w^2$ to the \sosw. The solution is obviously that a
fractional fill should instead contribute $f_iw^2$ to the \sosw, giving
the result $\sum f_iw^2=w^2$ for the $n$ fills, which is what we want.

Now we look at the case where we have $N$ sub-events in an event group
which are fully correlated as in a NLO calculation with a real
correction event and a number of counter-events. Let's assume
that we are measuring jet transverse momentum and we have one fill per
jet for $M$ jets. We have one weight per sub-event, $w_i$, and we
apply smearing such that each jet, $j$, is filled with a fraction
$\epsilon_{ji}$ in one bin and $1-\epsilon_{ji}$ in an adjacent
bin. Clearly, if these bins were joined, we would like to have
\begin{equation}
  \label{eq:0}
  \sow =\sum_{j,i} w_i \quad \text{and}\quad
  \sosw =\sum_j\left(\sum_iw_i\right)^2
\end{equation}
As before \sow is no problem, we simply fill
\begin{equation}
  \label{eq:1}
  \sum_j\sum_i\epsilon_{ji}w_i +
  \sum_j\sum_i(1-\epsilon_{ji})w_i = \sum_{j,i} w_i.
\end{equation}
For the \sosw we clearly cannot fill
\begin{equation}
  \label{eq:2}
  \sum_j\sum_i\epsilon_{ji}w^2_i +
  \sum_j\sum_i(1-\epsilon_{ji})w^2_i
\end{equation}
since we need to combine all the fills in the event group before
filling. Nor can we do
\begin{equation}
  \label{eq:3}
  \sum_j\left(\sum_i\epsilon_{ji}w_i\right)^2 +
  \sum_j\left(\sum_i(1-\epsilon_{ji})w_i\right)^2
\end{equation}
since we need to take into account the fractional fills. The trick is
now to realise that for a given jet, $j$, the NLO calculation requires
that the fill values are close to each other so also the
$\epsilon_{ji}$ are close to each other. We can therefore replace them
with the average $\bar{\epsilon}_j=\sum_i\epsilon_{ji}/N$ and fill \sosw with
\begin{equation}
  \label{eq:4}
  \sum_j\bar{\epsilon}_j\left(\sum_iw_i\right)^2 +
  \sum_j(1-\bar{\epsilon}_j)\left(\sum_iw_i\right)^2 =
  \sum_j\left(\sum_iw_i\right)^2
\end{equation}
which gives the correct \sosw. However, the averaging of
$\epsilon_{ij}$ means that we are assuming that the NLO calculation is
stable, and the errors will be calculated as if it was stable. We
would like to have a procedure where the errors becomes large if the
NLO calculation is unstable. Therefore we need a procedure which takes
into account that the $\epsilon_{ij}$ are not exactly the same for a
given $j$, while it still gives the result above if they are.

For two sub-events, the procedure should be such that if
$\epsilon_{1j}<\epsilon_{2j}$ we should have two fractional fills, one
filling $w_1+w_2$ with fraction $\epsilon_{1j}$, and one filling only
$w_1$ with fraction $\epsilon_{2j}-\epsilon_{1j}$ (and the
corresponding fills in the neighbouring bin).

For more sub-events, it becomes a bit more complicated, and even more
complicated if it is not obvious which jet in one sub-event
corresponds to which in another.

\rivet~3 implements a procedure defining a rectangular window of width
$\delta_{ij}$ around each fill point, $x_{ij}$. This width should be
smaller than the width of the corresponding histogram bin, and as
$x_{ij}\rightarrow x_{ik}$ we should have a smooth transition
$\delta_{ij}\rightarrow \delta_{ik}$. As an example we could use a
weighted average of the width of the bin corresponding to $x_{ij}$ and
the width of the closest neighbouring bin. So if we have bin edges
$b_k$ and the bin corresponding to $x_{ij}$ is between $b_k$ and
$b_{k-1}$ centred around $c_k=(b_k+b_{k-1})/2$ we would take
\begin{equation}
  \label{eq:5}
  \delta_{ij} = \left\{\begin{array}{ll}
  x_{ij} > c_k: \varepsilon\left(2\left(b_k-x_{ij}\right)+
    \frac{x_{ij}-c_{k}}{b_k-b_{k-1}}\left(b_{k+1}-b_{k-1}\right)\right)\\
  x_{ij} < c_k: \varepsilon\left(2\left(x_{ij}-b_{k-1}\right)+
    \frac{c_k-x_{ij}}{b_k-b_{k-1}}\left(b_{k}-b_{k-2}\right)\right)
\end{array}\right.,
\end{equation}
with $\varepsilon<1$.

\noindent
The procedure in \rivet~3 is therefore the following:
\begin{itemize}\itemsep 0mm
\item Collect all fills, $x_{ij}$ in all $N$ sub-events with weight
  $w_i$.
\item Construct the smearing windows $\delta_{ij}$.
\item Set the fill fraction to $f_{ij}=1/N$ (since we want each jet to
  sum up to one fill).
\item Construct all possible sub-windows from the edges,
  $x_{ij}\pm\delta_{ij}/2$, of all windows.
\item For each sub-window, $l$, with width $\delta_l$, sum up all fills
  which are overlapping (they will either completely overlap or not at
  all) and sum up
  \begin{eqnarray}
    \label{eq:6}
    w_l&=&\sum_{\delta_l\in\delta_{ij}}w_i\\
    f_l&=&\sum_{\delta_l\in\delta_{ij}}f_{ij}\delta_l/\delta_{ij}
  \end{eqnarray}
  and fill the histogram bin corresponding to the midpoint of the
  sub-window with weight $w_l$ and fraction $f_l$.
\end{itemize}

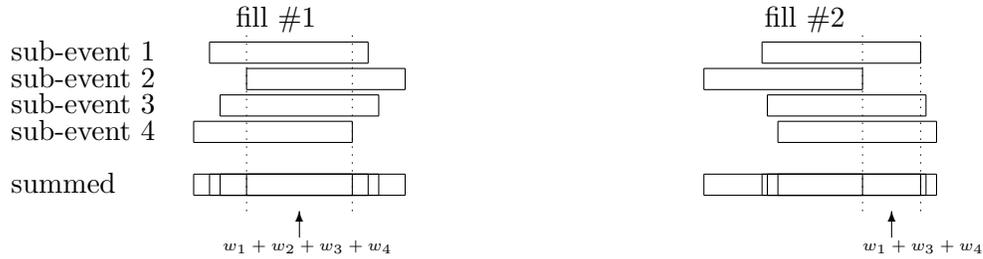
\begin{figure}
  \begin{center}
    \begin{picture}(300,60)
      \newsavebox{\fillwin}
      \savebox{\fillwin}(60,10)[bl]{
        \put(0,1){\line(1,0){60}}
        \put(0,9){\line(1,0){60}}
        \put(0,1){\line(0,1){8}}
        \put(60,1){\line(0,1){8}}}

      \put(25,50){\usebox{\fillwin}}
      \put(25,0){\usebox{\fillwin}}
      \put(234,50){\usebox{\fillwin}}
      \put(234,0){\usebox{\fillwin}}

      \put(39,40){\usebox{\fillwin}}
      \put(39,0){\usebox{\fillwin}}
      \put(212,40){\usebox{\fillwin}}
      \put(212,0){\usebox{\fillwin}}

      \put(29,30){\usebox{\fillwin}}
      \put(29,0){\usebox{\fillwin}}
      \put(236,30){\usebox{\fillwin}}
      \put(236,0){\usebox{\fillwin}}

      \put(19,20){\usebox{\fillwin}}
      \put(19,0){\usebox{\fillwin}}
      \put(240,20){\usebox{\fillwin}}
      \put(240,0){\usebox{\fillwin}}

      \put(-50,52){sub-event 1}
      \put(-50,42){sub-event 2}
      \put(-50,32){sub-event 3}
      \put(-50,22){sub-event 4}
      \put(-50,2){summed}

      \put(35,65){fill \#1}
      \put(235,65){fill \#2}

      \multiput(39,-5)(0,3){23}{\line(0,1){0.5}}
      \multiput(79,-5)(0,3){23}{\line(0,1){0.5}}
      \multiput(272,-5)(0,3){23}{\line(0,1){0.5}}
      \multiput(294,-5)(0,3){23}{\line(0,1){0.5}}

      \put(59,-15){\vector(0,1){10}}
      \put(30,-20){{\tiny $w_1+w_2+w_3+w_4$}}

      \put(283,-15){\vector(0,1){10}}
      \put(272,-20){{\tiny $w_1+w_3+w_4$}}

    \end{picture}
  \end{center}
  \caption{\label{fig:nlofill} A pictorial view of how the sub-windows
    are constructed for a one dimensional histogram.}

\end{figure}

\noindent
Figure~\ref{fig:nlofill} provides a pictorial illustration of how the
sub-windows are constructed for a one-dimensional histogram. This
procedure is easily extended to two-dimensional histograms, which will
done in a future release.

\end{appendix}

\bibliography{custom,auto}

\begin{thebibliography}{100}
\providecommand{\url}[1]{\texttt{#1}}
\providecommand{\urlprefix}{URL }
\expandafter\ifx\csname urlstyle\endcsname\relax
  \providecommand{\doi}[1]{doi:\discretionary{}{}{}#1}\else
  \providecommand{\doi}{doi:\discretionary{}{}{}\begingroup
  \urlstyle{rm}\Url}\fi
\providecommand{\eprint}[2][]{\url{#2}}

\bibitem{Buckley:2010ar}
A.~Buckley, J.~Butterworth, L.~Lönnblad, D.~Grellscheid, H.~Hoeth, J.~Monk,
  H.~Schulz and F.~Siegert,
\newblock \emph{{Rivet user manual}},
\newblock Comput. Phys. Commun. \textbf{184}, 2803 (2013),
\newblock \doi{10.1016/j.cpc.2013.05.021},
\newblock \eprint{1003.0694}.

\bibitem{Sjostrand:2014zea}
T.~Sjöstrand, S.~Ask, J.~R. Christiansen, R.~Corke, N.~Desai, P.~Ilten,
  S.~Mrenna, S.~Prestel, C.~O. Rasmussen and P.~Z. Skands,
\newblock \emph{{An Introduction to PYTHIA 8.2}},
\newblock Comput. Phys. Commun. \textbf{191}, 159 (2015),
\newblock \doi{10.1016/j.cpc.2015.01.024},
\newblock \eprint{1410.3012}.

\bibitem{Bellm:2015jjp}
J.~Bellm \emph{et~al.},
\newblock \emph{{Herwig 7.0/Herwig++ 3.0 release note}},
\newblock Eur. Phys. J. \textbf{C76}(4), 196 (2016),
\newblock \doi{10.1140/epjc/s10052-016-4018-8},
\newblock \eprint{1512.01178}.

\bibitem{Hoeche:2012yf}
S.~Hoeche, F.~Krauss, M.~Schonherr and F.~Siegert,
\newblock \emph{{QCD matrix elements + parton showers: The NLO case}},
\newblock JHEP \textbf{04}, 027 (2013),
\newblock \doi{10.1007/JHEP04(2013)027},
\newblock \eprint{1207.5030}.

\bibitem{Biro:2019ijx}
G.~Bíró, G.~G. Barnaföldi, G.~Papp, M.~Gyulassy, P.~Lévai, X.-N. Wang and
  B.-W. Zhang,
\newblock \emph{{Introducing HIJING++: the Heavy Ion Monte Carlo Generator for
  the High-Luminosity LHC Era}},
\newblock PoS \textbf{HardProbes2018}, 045 (2019),
\newblock \doi{10.22323/1.345.0045},
\newblock \eprint{1901.04220}.

\bibitem{Krauss:2018djz}
F.~Krauss, J.~M. Lindert, R.~Linten and M.~Schönherr,
\newblock \emph{{Accurate simulation of $W$, $Z$ and Higgs boson decays in
  Sherpa}},
\newblock Eur. Phys. J. \textbf{C79}(2), 143 (2019),
\newblock \doi{10.1140/epjc/s10052-019-6614-x},
\newblock \eprint{1809.10650}.

\bibitem{Hoche:2018gti}
S.~Höche, S.~Kuttimalai and Y.~Li,
\newblock \emph{{Hadronic Final States in DIS at NNLO QCD with Parton
  Showers}},
\newblock Phys. Rev. \textbf{D98}(11), 114013 (2018),
\newblock \doi{10.1103/PhysRevD.98.114013},
\newblock \eprint{1809.04192}.

\bibitem{Hoeche:2011fd}
S.~Hoeche, F.~Krauss, M.~Schonherr and F.~Siegert,
\newblock \emph{{A critical appraisal of NLO+PS matching methods}},
\newblock JHEP \textbf{09}, 049 (2012),
\newblock \doi{10.1007/JHEP09(2012)049},
\newblock \eprint{1111.1220}.

\bibitem{Buckley:2016bhy}
A.~Buckley and D.~Bakshi~Gupta,
\newblock \emph{{Powheg-Pythia matching scheme effects in NLO simulation of
  dijet events}}  (2016),
\newblock \eprint{1608.03577}.

\bibitem{CMS:2016kle}
{CMS Collaboration},
\newblock \emph{{Investigations of the impact of the parton shower tuning in
  Pythia 8 in the modelling of $\mathrm{t\overline{t}}$ at $\sqrt{s}=8$ and 13
  TeV}}  (2016).

\bibitem{Cooper:2011gk}
B.~Cooper, J.~Katzy, M.~L. Mangano, A.~Messina, L.~Mijovic and P.~Skands,
\newblock \emph{{Importance of a consistent choice of alpha(s) in the matching
  of AlpGen and Pythia}},
\newblock Eur. Phys. J. \textbf{C72}, 2078 (2012),
\newblock \doi{10.1140/epjc/s10052-012-2078-y},
\newblock \eprint{1109.5295}.

\bibitem{Heinemeyer:2013tqa}
{LHC Higgs Cross Section Working Group},
\newblock \emph{{Handbook of LHC Higgs Cross Sections: 3. Higgs Properties}}
  (2013),
\newblock \doi{10.5170/CERN-2013-004},
\newblock \eprint{1307.1347}.

\bibitem{Pierog:2013ria}
T.~Pierog, I.~Karpenko, J.~M. Katzy, E.~Yatsenko and K.~Werner,
\newblock \emph{{EPOS LHC: Test of collective hadronization with data measured
  at the CERN Large Hadron Collider}},
\newblock Phys. Rev. \textbf{C92}(3), 034906 (2015),
\newblock \doi{10.1103/PhysRevC.92.034906},
\newblock \eprint{1306.0121}.

\bibitem{Ilten:2014wma}
P.~Ilten,
\newblock \emph{{Soft QCD Measurements at LHCb}},
\newblock In \emph{{Proceedings, 2nd Conference on Large Hadron Collider
  Physics Conference (LHCP 2014): New York, USA, June 2-7, 2014}} (2014),
  \eprint{1409.0434}.

\bibitem{Helenius:2019gbd}
I.~Helenius and C.~O. Rasmussen,
\newblock \emph{{Hard diffraction in photoproduction with Pythia 8}},
\newblock Eur. Phys. J. \textbf{C79}(5), 413 (2019),
\newblock \doi{10.1140/epjc/s10052-019-6914-1},
\newblock \eprint{1901.05261}.

\bibitem{Cormier:2018tog}
K.~Cormier, S.~Plätzer, C.~Reuschle, P.~Richardson and S.~Webster,
\newblock \emph{{Parton showers and matching uncertainties in top quark pair
  production with Herwig 7}},
\newblock Eur. Phys. J. \textbf{C79}(11), 915 (2019),
\newblock \doi{10.1140/epjc/s10052-019-7370-7},
\newblock \eprint{1810.06493}.

\bibitem{Skands:2014pea}
P.~Skands, S.~Carrazza and J.~Rojo,
\newblock \emph{{Tuning PYTHIA 8.1: the Monash 2013 Tune}},
\newblock Eur. Phys. J. \textbf{C74}(8), 3024 (2014),
\newblock \doi{10.1140/epjc/s10052-014-3024-y},
\newblock \eprint{1404.5630}.

\bibitem{Khachatryan:2015pea}
{CMS Collaboration},
\newblock \emph{{Event generator tunes obtained from underlying event and
  multiparton scattering measurements}},
\newblock Eur. Phys. J. \textbf{C76}(3), 155 (2016),
\newblock \doi{10.1140/epjc/s10052-016-3988-x},
\newblock \eprint{1512.00815}.

\bibitem{ATLAS:2012uec}
{ATLAS Collaboration},
\newblock \emph{{Summary of ATLAS Pythia 8 tunes}}  (2012).

\bibitem{Ball:2014uwa}
R.~D. Ball \emph{et~al.},
\newblock \emph{{Parton distributions for the LHC Run II}},
\newblock JHEP \textbf{04}, 040 (2015),
\newblock \doi{10.1007/JHEP04(2015)040},
\newblock \eprint{1410.8849}.

\bibitem{Ball:2017nwa}
R.~D. Ball \emph{et~al.},
\newblock \emph{{Parton distributions from high-precision collider data}},
\newblock Eur. Phys. J. \textbf{C77}(10), 663 (2017),
\newblock \doi{10.1140/epjc/s10052-017-5199-5},
\newblock \eprint{1706.00428}.

\bibitem{Buckley:2009bj}
A.~Buckley, H.~Hoeth, H.~Lacker, H.~Schulz and J.~E. von Seggern,
\newblock \emph{{Systematic event generator tuning for the LHC}},
\newblock Eur. Phys. J. \textbf{C65}, 331 (2010),
\newblock \doi{10.1140/epjc/s10052-009-1196-7},
\newblock \eprint{0907.2973}.

\bibitem{Aad:2010ac}
{ATLAS Collaboration},
\newblock \emph{{Charged-particle multiplicities in pp interactions measured
  with the ATLAS detector at the LHC}},
\newblock New J. Phys. \textbf{13}, 053033 (2011),
\newblock \doi{10.1088/1367-2630/13/5/053033},
\newblock \eprint{1012.5104}.

\bibitem{Adam:2015qaa}
J.~Adam \emph{et~al.},
\newblock \emph{{Measurement of pion, kaon and proton production in
  proton–proton collisions at $\sqrt{s} = 7$ TeV}},
\newblock Eur. Phys. J. \textbf{C75}(5), 226 (2015),
\newblock \doi{10.1140/epjc/s10052-015-3422-9},
\newblock \eprint{1504.00024}.

\bibitem{Aad:2015eia}
{ATLAS Collaboration},
\newblock \emph{{Differential top-antitop cross-section measurements as a
  function of observables constructed from final-state particles using pp
  collisions at $\sqrt{s}=7$ TeV in the ATLAS detector}},
\newblock JHEP \textbf{06}, 100 (2015),
\newblock \doi{10.1007/JHEP06(2015)100},
\newblock \eprint{1502.05923}.

\bibitem{Khachatryan:2014uva}
{ATLAS Collboration},
\newblock \emph{{Differential Cross Section Measurements for the Production of
  a W Boson in Association with Jets in Proton–Proton Collisions at $\sqrt
  s=7$ TeV}},
\newblock Phys. Lett. \textbf{B741}, 12 (2015),
\newblock \doi{10.1016/j.physletb.2014.12.003},
\newblock \eprint{1406.7533}.

\bibitem{Sirunyan:2018ptc}
{CMS Collboration},
\newblock \emph{{Measurements of differential cross sections of top quark pair
  production as a function of kinematic event variables in proton-proton
  collisions at $ \sqrt{s}=13 $ TeV}},
\newblock JHEP \textbf{06}, 002 (2018),
\newblock \doi{10.1007/JHEP06(2018)002},
\newblock \eprint{1803.03991}.

\bibitem{Azzi:2019yne}
P.~Azzi \emph{et~al.},
\newblock \emph{{Report from Working Group 1}},
\newblock CERN Yellow Rep. Monogr. \textbf{7}, 1 (2019),
\newblock \doi{10.23731/CYRM-2019-007.1},
\newblock \eprint{1902.04070}.

\bibitem{Nejad:2016bci}
B.~Chokoufé~Nejad, W.~Kilian, J.~M. Lindert, S.~Pozzorini, J.~Reuter and
  C.~Weiss,
\newblock \emph{{NLO QCD predictions for off-shell $ t\overline{t} $ and $
  t\overline{t}H $ production and decay at a linear collider}},
\newblock JHEP \textbf{12}, 075 (2016),
\newblock \doi{10.1007/JHEP12(2016)075},
\newblock \eprint{1609.03390}.

\bibitem{Chala:2018qdf}
M.~Chala, R.~Gröber and M.~Spannowsky,
\newblock \emph{{Searches for vector-like quarks at future colliders and
  implications for composite Higgs models with dark matter}},
\newblock JHEP \textbf{03}, 040 (2018),
\newblock \doi{10.1007/JHEP03(2018)040},
\newblock \eprint{1801.06537}.

\bibitem{Bothmann:2016loj}
E.~Bothmann, P.~Ferrarese, F.~Krauss, S.~Kuttimalai, S.~Schumann and
  J.~Thompson,
\newblock \emph{{Aspects of perturbative QCD at a 100 TeV future hadron
  collider}},
\newblock Phys. Rev. \textbf{D94}(3), 034007 (2016),
\newblock \doi{10.1103/PhysRevD.94.034007},
\newblock \eprint{1605.00617}.

\bibitem{Kasieczka:2019dbj}
A.~Butter \emph{et~al.},
\newblock \emph{{The Machine Learning Landscape of Top Taggers}},
\newblock SciPost Phys. \textbf{7}, 014 (2019),
\newblock \doi{10.21468/SciPostPhys.7.1.014},
\newblock \eprint{1902.09914}.

\bibitem{Monk:2018clo}
P.~Hansen, J.~W. Monk and C.~Wiglesworth,
\newblock \emph{{A Wavelet Based Pile-Up Mitigation Method for the LHC
  Upgrade}}  (2018),
\newblock \eprint{1812.07412}.

\bibitem{Moore:2018lsr}
L.~Moore, K.~Nordström, S.~Varma and M.~Fairbairn,
\newblock \emph{{Reports of My Demise Are Greatly Exaggerated: $N$-subjettiness
  Taggers Take On Jet Images}},
\newblock SciPost Phys. \textbf{7}(3), 036 (2019),
\newblock \doi{10.21468/SciPostPhys.7.3.036},
\newblock \eprint{1807.04769}.

\bibitem{Altheimer:2013yza}
A.~Altheimer \emph{et~al.},
\newblock \emph{{Boosted Objects and Jet Substructure at the LHC. Report of
  BOOST2012, held at IFIC Valencia, 23rd-27th of July 2012}},
\newblock Eur. Phys. J. \textbf{C74}(3), 2792 (2014),
\newblock \doi{10.1140/epjc/s10052-014-2792-8},
\newblock \eprint{1311.2708}.

\bibitem{Brown:2019pzx}
S.~Brown, A.~Buckley, C.~Englert, J.~Ferrando, P.~Galler, D.~J. Miller,
  L.~Moore, M.~Russell, C.~White and N.~Warrack,
\newblock \emph{{TopFitter: Fitting top-quark Wilson Coefficients to Run II
  data}},
\newblock PoS \textbf{ICHEP2018}, 293 (2019),
\newblock \doi{10.22323/1.340.0293},
\newblock \eprint{1901.03164}.

\bibitem{Buckley:2015nca}
A.~Buckley, C.~Englert, J.~Ferrando, D.~J. Miller, L.~Moore, M.~Russell and
  C.~D. White,
\newblock \emph{{Global fit of top quark effective theory to data}},
\newblock Phys. Rev. \textbf{D92}(9), 091501 (2015),
\newblock \doi{10.1103/PhysRevD.92.091501},
\newblock \eprint{1506.08845}.

\bibitem{Neill:2018wtk}
D.~Neill, A.~Papaefstathiou, W.~J. Waalewijn and L.~Zoppi,
\newblock \emph{{Phenomenology with a recoil-free jet axis: TMD fragmentation
  and the jet shape}},
\newblock JHEP \textbf{01}, 067 (2019),
\newblock \doi{10.1007/JHEP01(2019)067},
\newblock \eprint{1810.12915}.

\bibitem{Reyer:2019obz}
M.~Reyer, M.~Schönherr and S.~Schumann,
\newblock \emph{{Full NLO corrections to 3-jet production and $\mathbf
  {R_{32}}$ at the LHC}},
\newblock Eur. Phys. J. \textbf{C79}(4), 321 (2019),
\newblock \doi{10.1140/epjc/s10052-019-6815-3},
\newblock \eprint{1902.01763}.

\bibitem{Bothmann:2018trh}
E.~Bothmann and L.~Debbio,
\newblock \emph{{Reweighting a parton shower using a neural network: the
  final-state case}},
\newblock JHEP \textbf{01}, 033 (2019),
\newblock \doi{10.1007/JHEP01(2019)033},
\newblock \eprint{1808.07802}.

\bibitem{Bendavid:2018nar}
\emph{{Les Houches 2017: Physics at TeV Colliders Standard Model Working Group
  Report}} (2018), \eprint{1803.07977}.

\bibitem{deFlorian:2016spz}
D.~de~Florian \emph{et~al.},
\newblock \emph{{Handbook of LHC Higgs Cross Sections: 4. Deciphering the
  Nature of the Higgs Sector}}  (2016),
\newblock \doi{10.2172/1345634, 10.23731/CYRM-2017-002},
\newblock \eprint{1610.07922}.

\bibitem{Papucci:2011wy}
M.~Papucci, J.~T. Ruderman and A.~Weiler,
\newblock \emph{{Natural SUSY Endures}},
\newblock JHEP \textbf{09}, 035 (2012),
\newblock \doi{10.1007/JHEP09(2012)035},
\newblock \eprint{1110.6926}.

\bibitem{Amrith:2018yfb}
S.~Amrith, J.~M. Butterworth, F.~F. Deppisch, W.~Liu, A.~Varma and D.~Yallup,
\newblock \emph{{LHC constraints on a $B-L$ gauge model using Contur}},
\newblock JHEP \textbf{05}, 154 (2019),
\newblock \doi{10.1007/JHEP05(2019)154},
\newblock \eprint{1811.11452}.

\bibitem{Brooijmans:2018xbu}
G.~Brooijmans \emph{et~al.},
\newblock \emph{{Les Houches 2017: Physics at TeV Colliders New Physics Working
  Group Report}},
\newblock In \emph{{Les Houches 2017: Physics at TeV Colliders Standard Model
  Working Group Report}} (2018), \eprint{1803.10379}.

\bibitem{Butterworth:2019iff}
J.~M. Butterworth, M.~Chala, C.~Englert, M.~Spannowsky and A.~Titov,
\newblock \emph{{Higgs phenomenology as a probe of sterile neutrinos}}  (2019),
\newblock \eprint{1909.04665}.

\bibitem{Butterworth:2016sqg}
J.~M. Butterworth, D.~Grellscheid, M.~Krämer, B.~Sarrazin and D.~Yallup,
\newblock \emph{{Constraining new physics with collider measurements of
  Standard Model signatures}},
\newblock JHEP \textbf{03}, 078 (2017),
\newblock \doi{10.1007/JHEP03(2017)078},
\newblock \eprint{1606.05296}.

\bibitem{Dobbs:2001ck}
M.~Dobbs and J.~B. Hansen,
\newblock \emph{{The HepMC C++ Monte Carlo event record for High Energy
  Physics}},
\newblock Comput. Phys. Commun. \textbf{134}, 41 (2001),
\newblock \doi{10.1016/S0010-4655(00)00189-2}.

\bibitem{Buckley:2019xhk}
A.~Buckley, P.~Ilten, D.~Konstantinov, L.~Lönnblad, J.~Monk, W.~Porkorski,
  T.~Przedzinski and A.~Verbytskyi,
\newblock \emph{{The HepMC3 Event Record Library for Monte Carlo Event
  Generators}}  (2019),
\newblock \eprint{1912.08005}.

\bibitem{Bjorken:1969wi}
J.~D. Bjorken and S.~J. Brodsky,
\newblock \emph{{Statistical model for electron-positron annihilation into
  hadrons}},
\newblock Phys. Rev. \textbf{D1}, 1416 (1970),
\newblock \doi{10.1103/PhysRevD.1.1416}.

\bibitem{Catani:1996jh}
S.~Catani and M.~H. Seymour,
\newblock \emph{{The Dipole formalism for the calculation of QCD jet
  cross-sections at next-to-leading order}},
\newblock Phys. Lett. \textbf{B378}, 287 (1996),
\newblock \doi{10.1016/0370-2693(96)00425-X},
\newblock \eprint{hep-ph/9602277}.

\bibitem{Adriani:2015iwv}
O.~Adriani \emph{et~al.},
\newblock \emph{{Measurements of longitudinal and transverse momentum
  distributions for neutral pions in the forward-rapidity region with the LHCf
  detector}},
\newblock Phys. Rev. \textbf{D94}(3), 032007 (2016),
\newblock \doi{10.1103/PhysRevD.94.032007},
\newblock \eprint{1507.08764}.

\bibitem{rivethi}
C.~Bierlich \emph{et~al.},
\newblock \emph{Confronting experimental data with heavy-ion models: \rivet for
  heavy ions},
\newblock In preparation (2019/20).

\bibitem{Bilandzic:2010jr}
A.~Bilandzic, R.~Snellings and S.~Voloshin,
\newblock \emph{{Flow analysis with cumulants: Direct calculations}},
\newblock Phys. Rev. \textbf{C83}, 044913 (2011),
\newblock \doi{10.1103/PhysRevC.83.044913},
\newblock \eprint{1010.0233}.

\bibitem{Bilandzic:2013kga}
A.~Bilandzic, C.~H. Christensen, K.~Gulbrandsen, A.~Hansen and Y.~Zhou,
\newblock \emph{{Generic framework for anisotropic flow analyses with
  multiparticle azimuthal correlations}},
\newblock Phys. Rev. \textbf{C89}(6), 064904 (2014),
\newblock \doi{10.1103/PhysRevC.89.064904},
\newblock \eprint{1312.3572}.

\bibitem{Aamodt:2010cz}
{ALICE Collboration},
\newblock \emph{{Centrality dependence of the charged-particle multiplicity
  density at mid-rapidity in Pb-Pb collisions at $\sqrt{s_{NN}}=2.76$ TeV}},
\newblock Phys. Rev. Lett. \textbf{106}, 032301 (2011),
\newblock \doi{10.1103/PhysRevLett.106.032301},
\newblock \eprint{1012.1657}.

\bibitem{Aamodt:2011vg}
{ALICE Collboration},
\newblock \emph{{Particle-yield modification in jet-like azimuthal di-hadron
  correlations in Pb-Pb collisions at $\sqrt{s_{NN}} = 2.76$ TeV}},
\newblock Phys. Rev. Lett. \textbf{108}, 092301 (2012),
\newblock \doi{10.1103/PhysRevLett.108.092301},
\newblock \eprint{1110.0121}.

\bibitem{Abelev:2012hxa}
{ALICE Collboration},
\newblock \emph{{Centrality Dependence of Charged Particle Production at Large
  Transverse Momentum in Pb--Pb Collisions at $\sqrt{s_{\rm{NN}}} = 2.76$
  TeV}},
\newblock Phys. Lett. \textbf{B720}, 52 (2013),
\newblock \doi{10.1016/j.physletb.2013.01.051},
\newblock \eprint{1208.2711}.

\bibitem{Abelev:2012wca}
{ALICE Collboration},
\newblock \emph{{Pion, Kaon, and Proton Production in Central Pb--Pb Collisions
  at $\sqrt{s_{NN}} = 2.76$ TeV}},
\newblock Phys. Rev. Lett. \textbf{109}, 252301 (2012),
\newblock \doi{10.1103/PhysRevLett.109.252301},
\newblock \eprint{1208.1974}.

\bibitem{Abbas:2013bpa}
{ALICE Collboration},
\newblock \emph{{Centrality dependence of the pseudorapidity density
  distribution for charged particles in Pb-Pb collisions at $\sqrt{s_{\rm NN}}$
  = 2.76 TeV}},
\newblock Phys. Lett. \textbf{B726}, 610 (2013),
\newblock \doi{10.1016/j.physletb.2013.09.022},
\newblock \eprint{1304.0347}.

\bibitem{ABELEV:2013zaa}
{ALICE Collboration},
\newblock \emph{{Multi-strange baryon production at mid-rapidity in Pb-Pb
  collisions at $\sqrt{s_{NN}}$ = 2.76 TeV}},
\newblock Phys. Lett. \textbf{B728}, 216 (2014),
\newblock \doi{10.1016/j.physletb.2014.05.052, 10.1016/j.physletb.2013.11.048},
\newblock [Erratum: Phys. Lett.B734,409(2014)],
\newblock \eprint{1307.5543}.

\bibitem{Abelev:2013haa}
{ALICE Collboration},
\newblock \emph{{Multiplicity Dependence of Pion, Kaon, Proton and Lambda
  Production in $p$-Pb Collisions at $\sqrt{s_\mathrm{NN}}$ = 5.02 TeV}},
\newblock Phys. Lett. \textbf{B728}, 25 (2014),
\newblock \doi{10.1016/j.physletb.2013.11.020},
\newblock \eprint{1307.6796}.

\bibitem{Adam:2015kda}
{ALICE Collboration},
\newblock \emph{{Centrality evolution of the charged-particle pseudorapidity
  density over a broad pseudorapidity range in Pb-Pb collisions at
  $\sqrt{s_{\rm NN}} =$ 2.76 TeV}},
\newblock Phys. Lett. \textbf{B754}, 373 (2016),
\newblock \doi{10.1016/j.physletb.2015.12.082},
\newblock \eprint{1509.07299}.

\bibitem{Adam:2016izf}
{ALICE Collboration},
\newblock \emph{{Anisotropic flow of charged particles in Pb-Pb collisions at
  $\sqrt{s_{\rm NN}}=5.02$ TeV}},
\newblock Phys. Rev. Lett. \textbf{116}(13), 132302 (2016),
\newblock \doi{10.1103/PhysRevLett.116.132302},
\newblock \eprint{1602.01119}.

\bibitem{Adam:2016ddh}
{ALICE Collboration},
\newblock \emph{{Centrality dependence of the pseudorapidity density
  distribution for charged particles in Pb-Pb collisions at $\sqrt{s_{\rm
  NN}}=5.02$ TeV}},
\newblock Phys. Lett. \textbf{B772}, 567 (2017),
\newblock \doi{10.1016/j.physletb.2017.07.017},
\newblock \eprint{1612.08966}.

\bibitem{Adam:2016iwf}
{ALICE Collboration},
\newblock \emph{{Insight into particle production mechanisms via angular
  correlations of identified particles in pp collisions at
  $\sqrt{\mathrm{s}}=7$ TeV}},
\newblock Eur. Phys. J. \textbf{C77}(8), 569 (2017),
\newblock \doi{10.1140/epjc/s10052-017-5129-6},
\newblock \eprint{1612.08975}.

\bibitem{Aad:2015wga}
{ATLAS Collboration},
\newblock \emph{{Measurement of charged-particle spectra in Pb+Pb collisions at
  $\sqrt{{s}_\mathsf{{NN}}} = 2.76$ TeV with the ATLAS detector at the LHC}},
\newblock JHEP \textbf{09}, 050 (2015),
\newblock \doi{10.1007/JHEP09(2015)050},
\newblock \eprint{1504.04337}.

\bibitem{Aad:2015zza}
{ATLAS Collboration},
\newblock \emph{{Measurement of the centrality dependence of the
  charged-particle pseudorapidity distribution in proton–lead collisions at
  $\sqrt{s_{_\text {NN}}} = 5.02$ TeV with the ATLAS detector}},
\newblock Eur. Phys. J. \textbf{C76}(4), 199 (2016),
\newblock \doi{10.1140/epjc/s10052-016-4002-3},
\newblock \eprint{1508.00848}.

\bibitem{Bearden:2004yx}
{BRAHMS Collboration},
\newblock \emph{{Charged meson rapidity distributions in central Au+Au
  collisions at $\sqrt{s_\mathrm{NN}} = 200$-GeV}},
\newblock Phys. Rev. Lett. \textbf{94}, 162301 (2005),
\newblock \doi{10.1103/PhysRevLett.94.162301},
\newblock \eprint{nucl-ex/0403050}.

\bibitem{Khachatryan:2016txc}
{CMS Collboration},
\newblock \emph{{Evidence for collectivity in pp collisions at the LHC}},
\newblock Phys. Lett. \textbf{B765}, 193 (2017),
\newblock \doi{10.1016/j.physletb.2016.12.009},
\newblock \eprint{1606.06198}.

\bibitem{Adamczyk:2016exq}
{STAR Collboration},
\newblock \emph{{Beam Energy Dependence of the Third Harmonic of Azimuthal
  Correlations in Au+Au Collisions at RHIC}},
\newblock Phys. Rev. Lett. \textbf{116}(11), 112302 (2016),
\newblock \doi{10.1103/PhysRevLett.116.112302},
\newblock \eprint{1601.01999}.

\bibitem{Bromley:1995np}
J.~Bromley, N.~Brook, A.~Bunyatyan, T.~Carli, G.~Grindhammer, M.~Kuhlen,
  R.~Mohr, M.~Hayes and L.~Lönnblad,
\newblock \emph{{HZTOOL: A package for Monte Carlo-data comparison at HERA
  (version 1.0)}},
\newblock In \emph{{Future physics at HERA. Proceedings, Workshop, Hamburg,
  Germany, September 25, 1995-May 31, 1996. Vol. 1, 2}} (1995).

\bibitem{Waugh:2006ip}
B.~M. Waugh, H.~Jung, A.~Buckley, L.~Lönnblad, J.~M. Butterworth and E.~Nurse,
\newblock \emph{{HZTool and Rivet: Toolkit and Framework for the Comparison of
  Simulated Final States and Data at Colliders}},
\newblock In \emph{{15th International Conference on Computing in High Energy
  and Nuclear Physics (CHEP 2006) Mumbai, Maharashtra, India, February 13-17,
  2006}} (2006), \eprint{hep-ph/0605034}.

\bibitem{RivetHZtool}
H.~Jung and S.~Pl{\"a}tzer,
\newblock \emph{{Rivet/HZTool}},
\newblock
  \urlprefix\url{https://indico.cern.ch/event/845653/contributions/3635913/attachments/1949581/3235689/RivetHZTool.pdf},
\newblock {Talk presented at \textit{MCEGs for future ep and eA facilities},
  Vienna} (2019).

\bibitem{Abt:1994ye}
{H1 Collaboration},
\newblock \emph{{Energy flow and charged particle spectrum in deep inelastic
  scattering at HERA}},
\newblock Z. Phys. \textbf{C63}, 377 (1994),
\newblock \doi{10.1007/BF01580319}.

\bibitem{Aid:1995we}
{H1 Collaboration},
\newblock \emph{{Transverse energy and forward jet production in the low-$x$
  regime at HERA}},
\newblock Phys. Lett. \textbf{B356}, 118 (1995),
\newblock \doi{10.1016/0370-2693(95)00804-T},
\newblock \eprint{hep-ex/9506012}.

\bibitem{Adloff:1999ws}
{H1 Collaboration},
\newblock \emph{{Measurements of transverse energy flow in deep inelastic
  scattering at HERA}},
\newblock Eur. Phys. J. \textbf{C12}, 595 (2000),
\newblock \doi{10.1007/s100520000287},
\newblock \eprint{hep-ex/9907027}.

\bibitem{Aktas:2007hn}
{H1 Collaboration},
\newblock \emph{{Tests of QCD factorisation in the diffractive production of
  dijets in deep-inelastic scattering and photoproduction at HERA}},
\newblock Eur. Phys. J. \textbf{C51}, 549 (2007),
\newblock \doi{10.1140/epjc/s10052-007-0325-4},
\newblock \eprint{hep-ex/0703022}.

\bibitem{Andreev:2015cwa}
{H1 Collaboration},
\newblock \emph{{Diffractive Dijet Production with a Leading Proton in $ep$
  Collisions at HERA}},
\newblock JHEP \textbf{05}, 056 (2015),
\newblock \doi{10.1007/JHEP05(2015)056},
\newblock \eprint{1502.01683}.

\bibitem{H1:2015dma}
{H1 and ZEUS Collaborations},
\newblock \emph{{Combination of differential D$^{*\pm}$ cross-section
  measurements in deep-inelastic ep scattering at HERA}},
\newblock JHEP \textbf{09}, 149 (2015),
\newblock \doi{10.1007/JHEP09(2015)149},
\newblock \eprint{1503.06042}.

\bibitem{Chekanov:2001bw}
{ZEUS Collaboration},
\newblock \emph{{Dijet photoproduction at HERA and the structure of the
  photon}},
\newblock Eur. Phys. J. \textbf{C23}, 615 (2002),
\newblock \doi{10.1007/s100520200936},
\newblock \eprint{hep-ex/0112029}.

\bibitem{Chekanov:2007rh}
{ZEUS Collaboration},
\newblock \emph{{Diffractive photoproduction of dijets in ep collisions at
  HERA}},
\newblock Eur. Phys. J. \textbf{C55}, 177 (2008),
\newblock \doi{10.1140/epjc/s10052-008-0598-2},
\newblock \eprint{0710.1498}.

\bibitem{Abramowicz:2012jz}
{ZEUS Collaboration},
\newblock \emph{{Inclusive-jet photoproduction at HERA and determination of
  alphas}},
\newblock Nucl. Phys. \textbf{B864}, 1 (2012),
\newblock \doi{10.1016/j.nuclphysb.2012.06.006},
\newblock \eprint{1205.6153}.

\bibitem{Athron:2018vxy}
P.~Athron \emph{et~al.},
\newblock \emph{{Combined collider constraints on neutralinos and charginos}},
\newblock Eur. Phys. J. \textbf{C79}(5), 395 (2019),
\newblock \doi{10.1140/epjc/s10052-019-6837-x},
\newblock \eprint{1809.02097}.

\bibitem{Aaboud:2017buf}
{ATLAS Collaboration},
\newblock \emph{{Measurement of detector-corrected observables sensitive to the
  anomalous production of events with jets and large missing transverse
  momentum in $pp$ collisions at $\mathbf {\sqrt{s}=13}$ TeV using the ATLAS
  detector}},
\newblock Eur. Phys. J. \textbf{C77}(11), 765 (2017),
\newblock \doi{10.1140/epjc/s10052-017-5315-6},
\newblock \eprint{1707.03263}.

\bibitem{Aaboud:2019jcc}
{ATLAS Collaboration},
\newblock \emph{{Searches for scalar leptoquarks and differential cross-section
  measurements in dilepton-dijet events in proton-proton collisions at a
  centre-of-mass energy of $\sqrt{s}$ = 13 TeV with the ATLAS experiment}},
\newblock Eur. Phys. J. \textbf{C79}(9), 733 (2019),
\newblock \doi{10.1140/epjc/s10052-019-7181-x},
\newblock \eprint{1902.00377}.

\bibitem{Buckley:2019stt}
A.~Buckley, D.~Kar and K.~Nordström,
\newblock \emph{{Fast simulation of detector effects in Rivet}}  (2019),
\newblock \eprint{1910.01637}.

\bibitem{Drees:2013wra}
M.~Drees, H.~Dreiner, D.~Schmeier, J.~Tattersall and J.~S. Kim,
\newblock \emph{{CheckMATE: Confronting your Favourite New Physics Model with
  LHC Data}},
\newblock Comput. Phys. Commun. \textbf{187}, 227 (2015),
\newblock \doi{10.1016/j.cpc.2014.10.018},
\newblock \eprint{1312.2591}.

\bibitem{Conte:2012fm}
E.~Conte, B.~Fuks and G.~Serret,
\newblock \emph{{MadAnalysis 5, A User-Friendly Framework for Collider
  Phenomenology}},
\newblock Comput. Phys. Commun. \textbf{184}, 222 (2013),
\newblock \doi{10.1016/j.cpc.2012.09.009},
\newblock \eprint{1206.1599}.

\bibitem{Balazs:2017moi}
C.~Balázs \emph{et~al.},
\newblock \emph{{ColliderBit: a GAMBIT module for the calculation of
  high-energy collider observables and likelihoods}},
\newblock Eur. Phys. J. \textbf{C77}(11), 795 (2017),
\newblock \doi{10.1140/epjc/s10052-017-5285-8},
\newblock \eprint{1705.07919}.

\bibitem{deFavereau:2013fsa}
J.~de~Favereau, C.~Delaere, P.~Demin, A.~Giammanco, V.~Lemaître, A.~Mertens
  and M.~Selvaggi,
\newblock \emph{{DELPHES 3, A modular framework for fast simulation of a
  generic collider experiment}},
\newblock JHEP \textbf{02}, 057 (2014),
\newblock \doi{10.1007/JHEP02(2014)057},
\newblock \eprint{1307.6346}.

\bibitem{ATLAS:2015uwa}
{ATLAS Collaboration},
\newblock \emph{{Data-driven determination of the energy scale and resolution
  of jets reconstructed in the ATLAS calorimeters using dijet and multijet
  events at $\sqrt{s\ }=8~TeV$}}  (2015).

\bibitem{ATL-PHYS-PUB-2015-041}
{The ATLAS Collaboration},
\newblock \emph{{Electron identification measurements in ATLAS using $\sqrt{s}$
  = 13 TeV data with 50 ns bunch spacing}},
\newblock \urlprefix\url{https://cds.cern.ch/record/2048202},
\newblock ATL-PHYS-PUB-2015-041 (2015).

\bibitem{ATLAS-CONF-2016-024}
{ATLAS Collaboration},
\newblock \emph{{Electron efficiency measurements with the ATLAS detector using
  the 2015 LHC proton-proton collision data}},
\newblock \urlprefix\url{https://cds.cern.ch/record/2157687},
\newblock ATLAS-CONF-2016-024 (2016).

\bibitem{Aaboud:2019ynx}
{ATLAS Collaboration},
\newblock \emph{{Electron reconstruction and identification in the ATLAS
  experiment using the 2015 and 2016 LHC proton-proton collision data at
  $\sqrt{s}$ = 13 TeV}},
\newblock Eur. Phys. J. \textbf{C79}(8), 639 (2019),
\newblock \doi{10.1140/epjc/s10052-019-7140-6},
\newblock \eprint{1902.04655}.

\bibitem{ATL-PHYS-PUB-2015-037}
{The ATLAS Collaboration},
\newblock \emph{{Muon reconstruction performance in early $\sqrt{s}$ = 13 TeV
  data}},
\newblock \urlprefix\url{http://cds.cern.ch/record/2047831},
\newblock ATL-PHYS-PUB-2015-037 (2015).

\bibitem{Aad:2016jkr}
{ATLAS Collaboration},
\newblock \emph{{Muon reconstruction performance of the ATLAS detector in
  proton–proton collision data at $\sqrt{s}$ =13 TeV}},
\newblock Eur. Phys. J. \textbf{C76}(5), 292 (2016),
\newblock \doi{10.1140/epjc/s10052-016-4120-y},
\newblock \eprint{1603.05598}.

\bibitem{Aad:2014rga}
{ATLAS Collaboration},
\newblock \emph{{Identification and energy calibration of hadronically decaying
  tau leptons with the ATLAS experiment in $pp$ collisions at $\sqrt{s}$=8
  TeV}},
\newblock Eur. Phys. J. \textbf{C75}(7), 303 (2015),
\newblock \doi{10.1140/epjc/s10052-015-3500-z},
\newblock \eprint{1412.7086}.

\bibitem{ATL-PHYS-PUB-2015-045}
{The ATLAS Collaboration},
\newblock \emph{{Reconstruction, Energy Calibration, and Identification of
  Hadronically Decaying Tau Leptons in the ATLAS Experiment for Run-2 of the
  LHC}},
\newblock ATL-PHYS-PUB-2015-045 (2015).

\bibitem{Aaboud:2016yuq}
{ATLAS Collaboration},
\newblock \emph{{Measurement of the photon identification efficiencies with the
  ATLAS detector using LHC Run-1 data}},
\newblock Eur. Phys. J. \textbf{C76}(12), 666 (2016),
\newblock \doi{10.1140/epjc/s10052-016-4507-9},
\newblock \eprint{1606.01813}.

\bibitem{ATL-PHYS-PUB-2016-014}
{The ATLAS Collaboration},
\newblock \emph{{Photon identification in 2015 ATLAS data}},
\newblock \urlprefix\url{http://cds.cern.ch/record/2203125},
\newblock ATL-PHYS-PUB-2016-014 (2016).

\bibitem{Aad:2012re}
{ATLAS Collaboration},
\newblock \emph{{Performance of Missing Transverse Momentum Reconstruction in
  Proton-Proton Collisions at 7 TeV with ATLAS}},
\newblock Eur. Phys. J. \textbf{C72}, 1844 (2012),
\newblock \doi{10.1140/epjc/s10052-011-1844-6},
\newblock \eprint{1108.5602}.

\bibitem{Khachatryan:2014gga}
{CMS Collaboration},
\newblock \emph{{Performance of the CMS missing transverse momentum
  reconstruction in $pp$ data at $\sqrt{s}$ = 8 TeV}},
\newblock JINST \textbf{10}(02), P02006 (2015),
\newblock \doi{10.1088/1748-0221/10/02/P02006},
\newblock \eprint{1411.0511}.

\bibitem{CMS-PAS-JME-17-001}
{CMS Collaboration},
\newblock \emph{{Performance of missing transverse momentum in $pp$ collisions
  at $\sqrt{s}=13$ TeV using the CMS detector}},
\newblock \urlprefix\url{https://cds.cern.ch/record/2666972},
\newblock CMS-PAS-JME-17-001 (2018).

\bibitem{Inspire}
{\textsc{Inspire} group},
\newblock \emph{{The INSPIRE project}},
\newblock \urlprefix\url{https://inspirehep.net/} (2010).

\bibitem{Gleisberg:2008ta}
T.~Gleisberg, S.~Hoeche, F.~Krauss, M.~Schonherr, S.~Schumann, F.~Siegert and
  J.~Winter,
\newblock \emph{{Event generation with SHERPA 1.1}},
\newblock JHEP \textbf{02}, 007 (2009),
\newblock \doi{10.1088/1126-6708/2009/02/007},
\newblock \eprint{0811.4622}.

\end{thebibliography}

\nolinenumbers

\end{document}